\documentclass[aps,prb,reprint,eqsecnum,nofootinbib,notitlepage,superscriptaddress,floatfix,showpacs]{revtex4-1}

\usepackage[dvipdfmx]{graphicx} %For KS's TeX editor
\usepackage{amsmath,amssymb,amsfonts}
\usepackage{bm,braket,array}
\usepackage{comment}
\usepackage{mathrsfs} %RSFS font

\usepackage[usenames]{color}

\renewcommand{\k}{{\bm k}}
\newcommand{\K}{{\bm K}}

\newcommand{\Z}{\mathbb{Z}}

\newcommand{\bk}{\bm{k}}

\def\widebar{\accentset{{\cc@style\underline{\mskip10mu}}}} %widebar
\def\wideubar{\underaccent{{\cc@style\underline{\mskip10mu}}}} %wideubar

\begin{document}

\title{M\"obius topological superconductivity in UPt$_3$}

\author{Youichi Yanase}
\email[]{yanase@scphys.kyoto-u.ac.jp}
\affiliation{Department of Physics, Graduate School of Science, Kyoto University, Kyoto 606-8502, Japan}
\author{Ken Shiozaki}
\affiliation{Department of Physics, University of Illinois at Urbana Champaign, Urbana, IL 61801, USA}

\date{\today}

\begin{abstract}
Intensive studies for more than three decades have elucidated multiple superconducting phases and 
odd-parity Cooper pairs in a heavy fermion superconductor UPt$_3$. 
We identify a time-reversal invariant superconducting phase of UPt$_3$ as a recently proposed 
topological nonsymmorphic superconductivity. Combining the band structure of UPt$_3$, 
order parameter of $E_{\rm 2u}$ representation allowed by $P6_3/mmc$ space group symmetry, 
and topological classification by $K$-theory, 
we demonstrate the nontrivial $\Z_2$-invariant of three-dimensional DIII class enriched by glide symmetry. 
Correspondingly, double Majorana cone surface states appear at the surface Brillouin zone boundary. 
Furthermore, we show a variety of surface states and clarify the topological protection 
by crystal symmetry. Majorana arcs corresponding to tunable Weyl points appear 
in the time-reversal symmetry broken B-phase. Majorana cone protected by mirror Chern number 
and Majorana flat band by glide-winding number are also revealed.
\end{abstract}

\maketitle

\section{Introduction}

Unconventional superconductivity in strongly correlated electron systems is attracting renewed interest 
because it may be a platform of topological superconductivity accompanied by Majorana surface/edge/vortex/end 
states~\cite{Qi-Zhang,Tanaka_review,Sato-Fujimoto_review}. 
Although previous studies focused on the proximity-induced topological superconductivity in $s$-wave superconductor (SC)
heterostructures~\cite{Lutchyn2010,Mourik,Nadj-Perge,Fu-Kane2008,Sun2016}, natural $s$-wave SCs are mostly trivial 
from the viewpoint of topology. 
On the other hand, unconventional SCs may have topologically nontrivial properties originating from 
non-$s$-wave Cooper pairing. In particular, time-reversal symmetry (TRS) broken chiral SCs~\cite{Read-Green} 
and odd-parity spin-triplet SCs~\cite{Kitaev2001,Schnyder,Sato2010} are known to be candidates of 
topological superconductivity.  
However, from the viewpoint of materials science, evidences for chiral and/or odd-parity superconductivity have been reported 
for only a few materials, such as URu$_2$Si$_2$~\cite{Kasahara2007,Yano2008,Kittaka2016}, SrPtAs~\cite{BiswasSrPtAs}, 
Sr$_2$RuO$_4$~\cite{Sr2RuO4_review2012}, Cu$_x$Bi$_2$Se$_3$~\cite{Sasaki2011}, and ferromagnetic heavy fermion 
SCs~\cite{Aoki_review}.

Superconductivity in UPt$_3$ has been discovered in 1980's~\cite{Stewart}. Multiple superconducting phases illustrated in 
Fig.~\ref{phasediagram}~\cite{Fisher,Bruls,Adenwalla,Tou_UPt3} unambiguously exhibit exotic Cooper pairing which is probably 
categorized into the two-dimensional (2D) irreducible representation of point group $D_{\rm 6h}$~\cite{Sigrist-Ueda}. 
After several theoretical proposals examined by experiments for more than three decades, the $E_{\rm 2u}$ representation 
has been regarded as the most reasonable symmetry of superconducting order parameter~\cite{Sauls,Joynt}. 
In particular, the multiple phase diagram in the temperature-magnetic field plane is naturally reproduced by assuming 
a weak symmetry breaking term of hexagonal symmetry~\cite{Sauls}. Furthermore, a phase-sensitive measurement~\cite{Strand} and 
the observation of spontaneous TRS breaking~\cite{Schemm} in the low-temperature and low-magnetic field B-phase, 
which was predicted in the $E_{\rm 2u}$-state, support the $E_{\rm 2u}$ symmetry of superconductivity. 
The order parameter of $E_{\rm 2u}$ symmetry represents odd-parity spin-triplet Cooper pairs. 
Therefore, topologically nontrivial superconductivity is expected in UPt$_3$.  

\begin{figure}[htbp]
\begin{center}
\includegraphics[width=65mm]{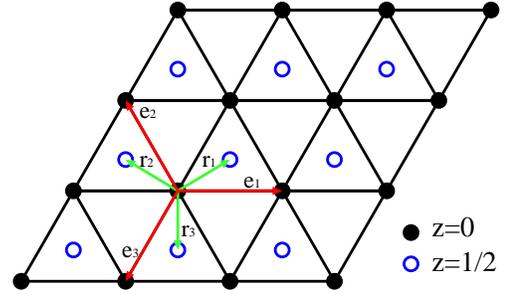}
\caption{(Color online)
Crystal structure of UPt$_3$. Uranium ions form AB stacked triangular lattice. 
2D vectors, ${\bm e}_i$ and ${\bm r}_i$, are shown by arrows. 
} 
\label{crystal}
\end{center}
\end{figure}

Furthermore, UPt$_3$ has an intriguing feature in the crystal structure, which is illustrated in Fig.~\ref{crystal}. 
The symmetry of the crystal is represented by nonsymmorphic space group $P6_{3}/mmc$~\cite{Comment2}; 
glide and screw symmetries including a half translation along the {\it c}-axis are preserved 
in spite of broken mirror and rotation symmetries. Exotic properties ensured by glide and/or screw symmetry 
are one of the central topics in the modern condensed matter physics. 
This topic for SCs traces back to Norman's work in 1995 for UPt$_3$~\cite{Norman1995}; 
a counterexample of Blount's theorem~\cite{Blount,Kobayashi-Sato}. 
The line nodal excitation predicted by Norman has been revisited by recent studies; group-theoretical 
proof~\cite{Micklitz2009,Micklitz2017-1}, microscopic origin~\cite{Yanase_UPt3_Weyl}, and 
topological protection~\cite{Kobayashi-Yanase-Sato} have been elucidated, and they have been confirmed 
by a first principles-based calculation~\cite{Nomoto}.

Recent developments in the theory of nonsymmorphic topological states of matter~\cite{Fang-Fu2015,Shiozaki2015,Shiozaki2016,Varjas2015,Liu2016} 
have uncovered novel topological insulators and SCs enriched by glide and/or screw symmetry, which are distinct from 
those classified by existing topological periodic table for symmorphic systems~\cite{Schnyder,Kitaev2009,Ryu2010,Morimoto2013,Shiozaki2014}. 
Since eigenvalues of glide and two-fold screw operators are $4\pi$-periodic, a M\"obius structure appears in the wave function 
and changes the topological classification.
Although such topological nonsymmorphic crystalline insulators have been proposed in KHgX (X = As, Sb, Bi)~\cite{KHgX,KHgX_ARPES} 
and CeNiSn~\cite{CeNiSn}, topological nonsymmorphic crystalline superconductor (TNSC) has not been identified in materials. 
In this paper we show the topological invariant specifying the TNSC by $K$-theory, and demonstrate its nontrivial value 
in UPt$_3$.

Multiband structures give rise to both intriguing and complicated aspects of many heavy fermion systems. 
However, the band structure of UPt$_3$ has been clarified to be rather simple~\cite{Joynt,Taillefer1988,Kimura_UPt3,McMullan}. 
Fermi surfaces (FSs) are classified into the two classes. 
The FSs of one class enclose the $A$-point in the Brillouin zone (BZ) [band 1 and band 2 in Ref.~\onlinecite{McMullan}], 
while those of the other class are centered on the $\Gamma$-point or $K$-point [bands 3, 4, and 5 
in Ref.~\onlinecite{McMullan}]. 
The two classes are not hybridized in the surface state on the (100)-direction where the glide symmetry is preserved. 
Therefore, we can separately study the topological invariants and surface states arising from the multiband FSs. 
The TNSC is attributed to the former FSs in Sec.~V. 
The latter FSs are also accompanied by various topological surface states, for which we identify topological invariant in Sec.~VI.

The paper is organized as follows. 
In Sec.~II, we introduce a minimal two-sublattice model for nonsymmorphic superconductivity in UPt$_3$. 
In Sec.~IIB, Dirac nodal lines protected by $P6_3/mmc$ space group symmetry are proved. 
In Sec.~IIC, the order parameter of $E_{\rm 2u}$ symmetry is explained. 
The calculated surface states on the glide invariant (100)-surface are shown in Sec.~III. 
In Sec.~IV, three-dimensional (3D) TNSC of DIII class is classified on the basis of the $K$-theory. 
In Sec.~V, we show that the glide-$\Z_2$ invariant characterizing the TNSC is nontrivial in UPt$_3$ A-phase. 
The underlying origin of the TNSC accompanied by double Majorana cone surface states is discussed. 
In Sec.~VI, we characterize other topological surface states by low-dimensional topological invariants 
enriched by crystal mirror, glide, and rotation symmetries. Constraints on these topological invariants by 
nonsymmorphic space group symmetry are also revealed. Finally, a brief summary is given in Sec.~VII. 
Ingredients giving rise to rich topological properties of UPt$_3$ are discussed.

\section{Model}\label{sec:model}

\subsection{Nonsymmorphic two-sublattice model}

We study the superconducting state in UPt$_3$ by analyzing the Bogoliubov-de Gennes (BdG) Hamiltonian 
for nonsymmorphic two-sublattice model~\cite{Yanase_UPt3_Weyl}, 
\begin{align}
{\cal H}_{\rm BdG}&= \sum_{{\bm k},m,s} \xi({\bm k}) c_{{\bm k}ms}^\dagger c_{{\bm k}ms} 
+ \sum_{{\bm k},s} \left[a({\bm k}) c^\dagger_{{\bm k}1s}c_{{\bm k}2s} + {\rm h.c.}\right]
\nonumber \\
& +\sum_{{\bm k},m,s,s'} \alpha_m {\bm g}({\bm k}) \cdot {\bm s}_{ss'}c^\dagger_{{\bm k}ms}c_{{\bm k}ms'}
\nonumber \\ 
& + \frac{1}{2} \sum_{{\bm k},m,m',s,s'} \left[\Delta_{mm'ss'}({\bm k}) c^\dagger_{{\bm k}ms}c^\dagger_{-{\bm k}m's'} + {\rm h.c}  \right], 
\label{eq:model}
\end{align}
where ${\bm k}$, $m=1,2$, and $s=\uparrow,\downarrow$ are index of momentum, sublattice, and spin, respectively. 
The last term represents the gap function and others are normal part Hamiltonian. 
Taking into account the crystal structure of UPt$_3$ illustrated in Fig.~\ref{crystal}, 
we adopt an intra-sublattice kinetic energy, 
\begin{align}
\xi({\bm k}) = 2 t \sum_{i=1,2,3}\cos{\bm k}_\parallel\cdot{\bm e}_i + 2 t_z \cos k_z -\mu, 
\end{align}
and an inter-sublattice hopping term,  
\begin{align}
a({\bm k}) = 2 t' \cos\frac{k_z}{2} \sum_{i=1,2,3} e^{i{\bm k}_\parallel\cdot{\bm r}_i}, 
\end{align}
with ${\bm k}_\parallel=(k_x,k_y)$. The basis translation vectors in two dimension are 
${\bm e}_1 = (1,0)$, ${\bm e}_2 = (-\frac{1}{2},\frac{\sqrt{3}}{2})$, and 
${\bm e}_3 = (-\frac{1}{2},-\frac{\sqrt{3}}{2})$. The interlayer neighboring vectors projected onto the 
basal plane are given by ${\bm r}_1 = (\frac{1}{2},\frac{1}{2\sqrt{3}})$, ${\bm r}_2 = (-\frac{1}{2},\frac{1}{2\sqrt{3}})$, 
and ${\bm r}_3 = (0,-\frac{1}{\sqrt{3}})$. These 2D vectors are illustrated in Fig.~\ref{crystal}. 

Although the crystal point group symmetry is centrosymmetric $D_{\rm 6h}$,  
local point group symmetry at Uranium ions is $D_{\rm 3h}$ lacking inversion symmetry. Then, Kane-Mele spin-orbit coupling 
(SOC)~\cite{Kane-Mele} with 
$g$-vector~\cite{Saito_MoS2} 
\begin{align}
{\bm g}({\bm k})= \hat{z} \sum_{i=1,2,3} \sin{\bm k}_\parallel\cdot{\bm e}_i, 
\label{g-vector}
\end{align}
is allowed by symmetry. The coupling constant has to be sublattice-dependent, 
$(\alpha_1,\alpha_2)=(\alpha,-\alpha)$, 
so as to preserve the global $D_{\rm 6h}$ point group symmetry~\cite{Kane-Mele,Fischer,JPSJ.81.034702}.

Quantum oscillation measurements combined with band structure 
calculations~\cite{Joynt,Taillefer1988,Kimura_UPt3,McMullan,Nomoto} have shown a pair of FSs
centered at the $A$-point ($A$-FSs) on the BZ face. 
Interestingly, the paired bands are degenerate on the $A$-$L$ lines and form 
Dirac nodal lines~\cite{Burkov-Hook-Balents2011}, 
which have been experimentally observed by de Haas-van Alphen experiments~\cite{McMullan}.
In the next subsection we show that the Dirac nodal lines are protected by the nonsymmorphic space group 
symmetry of $P6_3/mmc$ (No.~194)~\cite{Yanase_UPt3_Weyl,Kobayashi-Yanase-Sato}. 
Thus, the two A-FSs are naturally paired by the nonsymmorphic crystal symmetry. 
By choosing a parameter set $(t,t_z,t',\alpha,\mu)=(1,-4,1,2,12)$ our two band model reproduces 
the paired $A$-FSs. In this paper we show that the peculiar band structure 
results in exotic superconductivity in terms of symmetry and topology.

First principles band structure calculations also predict three FSs centered on the $\Gamma$-point ($\Gamma$-FSs), 
and two FSs enclosing the $K$-point ($K$-FSs)~\cite{Joynt,Taillefer1988,Nomoto}, 
although the existence of $K$-FSs is experimentally under debates~\cite{McMullan}. 
We show that a variety of topological surface states may arise from these bands. 
A parameter set $(t,t_z,t',\alpha,\mu)=(1,4,1,0,16)$ reproduces one of the $\Gamma$-FSs, while 
another set $(t,t_z,t',\alpha,\mu)=(1,-1,0.4,0.2,-5.2)$ is adopted for the $K$-FSs.

\subsection{Dirac nodal lines in space group $P6_3/mmc$}\label{sec:Dirac-node}

The single particle states are four-fold degenerate on the $A$-$L$ lines 
[$\k = (0, k_y, \pi)$ and symmetric lines]. 
In addition to the usual Kramers degeneracy, the sublattice degree of freedom gives additional degeneracy.
This feature is reproduced in the normal part Hamiltonian, because the inter-sublattice hopping vanishes 
on the BZ face ($k_z =\pi$) and the SOC disappears on the $A$-$L$ lines. 
Below we show that the existence of Dirac line nodes is ensured by the space group symmetry. 

First, we show the additional degeneracy in the absence of the SOC. 
In the SU(2) symmetric case, the two spin states are equivalent and naturally degenerate. 
Then, we can define the TRS, $T=K$, and screw symmetry $S^z_\pi(k_z)$ in each spin sector, 
where $K$ is the complex conjugate operator. 
At the BZ face, $k_z=\pi$, we have $S^z_\pi(\pi)=i\sigma_y$ where $\sigma_i$ is the Pauli matrix in the sublattice space. 
Because the combined magnetic-screw symmetry satisfies $\left[S^z_\pi(\pi)T\right]^2=-1$, 
the two-fold degeneracy in each spin sector is proved by familiar Kramers theorem. 
Taking into account the spin-degeneracy, we obtain four-fold degenerate bands on the entire BZ face.

The four-fold degeneracy is partly lifted by the SOC. However, the degeneracy of two spinful bands 
is protected on the $A$-$L$ lines, that is proved as follows.
The little group on the $A$-$L$ lines includes the rotation symmetry $IG^{xz}(k_z)$, mirror symmetry $M^{yz}$, 
and magnetic-inversion symmetry $IT$. 
We here represent $T=is_y K$, $I=\sigma_x$, and $M^{yz}= is_x$, respectively. 
The glide symmetry is represented by $G^{xz}(k_z)=s_y \sigma_y$ at $k_z=\pm \pi$ 
while $G^{xz}(k_z)=is_y \sigma_x$ at $k_z=0$. 
The nonsymmorphic property of rotation symmetry is emphasized by denoting as $IG^{xz}(k_z)$. 
The symmetry operations satisfy the algebra 
\begin{align}
&\left[IG^{xz}(\pi)\right]^2=-1, 
\label{AL_algebra1}
\\
&\{IG^{xz}(\pi),IT\}=0, 
\label{AL_algebra2}
\\
&\{IG^{xz}(\pi),M^{yz}\}=0.  
\label{AL_algebra3}
\end{align}
The first relation ensures the sector decomposition to the $\lambda=\pm i$ eigenstates of rotation operator. 
Because the $IT$ symmetry is preserved in each subsector, the Kramers theorem holds. 
The anti-commutation relation of two unitary symmetries, Eq.~(\ref{AL_algebra3}), shows that 
a Kramers pair in one subsector has to be degenerate with another Kramers pair in the other subsector. 
Therefore, the four-fold degeneracy on the $A$-$L$ lines is protected by symmetry.

\begin{figure}[htbp]
\begin{center}
\includegraphics[width=75mm]{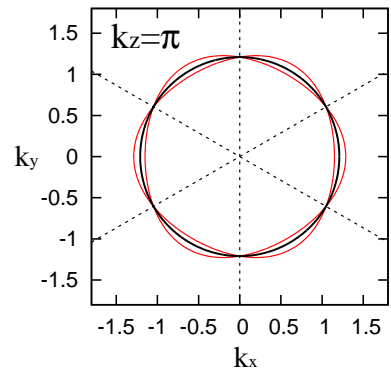}
\caption{(Color online)
FSs on the BZ face, $k_z=\pi$. Thin red lines show the FSs in the presence of the SOC ($\alpha=1$), 
while the thick black line is the overlapping FSs in the absence of the SOC ($\alpha=0$). 
Dashed lines show the $A$-$L$ lines in the first BZ. 
We set parameters $(t,t_z,t',\mu)=(1,-4,1,12)$ to reproduce the $A$-FSs. 
} 
\label{Dirac_node}
\end{center}
\end{figure}

Figure~\ref{Dirac_node} shows the FSs in our model. It is illustrated that the FSs completely overlap 
on the BZ face in the absence of the SOC. Although the SOC splits the FSs, the degeneracy remains on the $A$-$L$ lines. 
These features are consistent with above group theoretical analysis.

The Dirac nodal lines in the $P6_3/mmc$ space group are one of the typical examples of band degeneracy 
protected by nonsymmorphic crystal symmetry~\cite{Young2012,Watanabe-Po-Vishwanath,Niu2016,Yang2017,Bradley}. 
On the BZ face away from the $A$-$L$ lines, the non-Kramers degeneracy is lifted purely by the SOC. 
Therefore, the SOC gives particularly significant effects on the BZ face. 
This is the underlying origin of the SOC-induced nodal loop in the superconducting gap~\cite{Yanase_UPt3_Weyl,Kobayashi-Yanase-Sato}.

\subsection{Order parameter of $E_{\rm 2u}$ representation}\label{sec:order-parameter}

The multiple superconducting phases in UPt$_3$ have been reasonably attributed to two-component order parameters 
in the $E_{\rm 2u}$ irreducible representation of $D_{\rm 6h}$ point group~\cite{Sauls,Joynt}. 
The gap function is generally represented by  
\begin{align}
\hat{\Delta}({\bm k}) = \eta_1 \hat{\Gamma}^{E_{\rm 2u}}_1 + \eta_2 \hat{\Gamma}^{E_{\rm 2u}}_2. 
\end{align}
The two-component order parameters are parametrized as 
\begin{align}
(\eta_1, \eta_2) = \Delta (1,i \eta)/\sqrt{1+\eta^2}, 
\end{align}
with a real variable $\eta$. 
The basis functions $\hat{\Gamma}^{E_{\rm 2u}}_1$ and $\hat{\Gamma}^{E_{\rm 2u}}_2$ 
are admixture of some harmonics. 
Adopting the neighboring Cooper pairs in the crystal lattice of U ions, we obtain the basis functions 
\begin{align}
&
\hat{\Gamma}^{E_{\rm 2u}}_1 = \Bigl[\delta \left\{p_x({\bm k})s_x - p_y({\bm k})s_y\right\} \sigma_0 
\nonumber \\ & \hspace{8mm} 
+ f_{(x^2-y^2)z}({\bm k})s_z \sigma_x - d_{yz}({\bm k})s_z \sigma_y \Bigr] i s_y,
\label{Gamma_1}
\\
&
\hat{\Gamma}^{E_{\rm 2u}}_2 = \Bigl[\delta \left\{p_y({\bm k})s_x + p_x({\bm k})s_y\right\} \sigma_0 
\nonumber \\ & \hspace{8mm} 
+ f_{xyz}({\bm k})s_z \sigma_x - d_{xz}({\bm k})s_z \sigma_y \Bigr] i s_y, 
\label{Gamma_2}
\end{align}
which are composed of the $p$-wave, $d$-wave, and $f$-wave components given by 
\begin{eqnarray}
p_{x}({\bm k}) = \sum_{i} e_i^{x}  \sin {\bm k}_\parallel\cdot{\bm e}_i, 
\label{SM_p1}
\\
p_{y}({\bm k}) = \sum_{i} e_i^{y}  \sin {\bm k}_\parallel\cdot{\bm e}_i,  
\label{SM_p2}
\end{eqnarray}
\begin{eqnarray}
d_{xz}({\bm k}) = - \sqrt{3} \sin\frac{k_z}{2} {\rm Im} \sum_{i} r_i^{x} e^{i {\bm k}_\parallel \cdot {\bm r}_i}, 
\label{SM_d1}
\\
d_{yz}({\bm k}) = - \sqrt{3} \sin\frac{k_z}{2} {\rm Im} \sum_{i} r_i^{y} e^{i {\bm k}_\parallel \cdot {\bm r}_i}, 
\label{SM_d2}
\end{eqnarray}
\begin{eqnarray}
f_{xyz}({\bm k}) = -\sqrt{3} \sin\frac{k_z}{2} {\rm Re} \sum_{i} r_i^{x} e^{i {\bm k}_\parallel \cdot {\bm r}_i},  
\label{SM_f1}
\\
f_{(x^2-y^2)z}({\bm k}) = -\sqrt{3} \sin\frac{k_z}{2} {\rm Re} \sum_{i} r_i^{y} e^{i {\bm k}_\parallel \cdot {\bm r}_i}.  
\label{SM_f2}
\end{eqnarray} 
Pauli matrix in the spin and sublattice space are denoted by $s_i$ and $\sigma_i$, respectively. 

The purely $f$-wave state has been intensively investigated, 
and the phase diagram compatible with UPt$_3$ has been obtained~\cite{Sauls}. However, an admixture of 
a $p$-wave component is allowed by symmetry and it changes the gap structure and topological 
properties~\cite{Yanase_UPt3_Weyl}. 
Thus, we here take into account a small $p$-wave component with $0 < |\delta| \ll 1$. 
The small $p$-wave component does not alter the phase diagram consistent with experiments. 
On the other hand, the dominantly $p$-wave state discussed in Ref.~\onlinecite{Nomoto} would fail to reproduce 
the phase diagram.

Besides the $p$-wave component, a sublattice-singlet spin-triplet $d$-wave component accompanies 
the $f$-wave component as a result of the nonsymmorphic crystal structure of UPt$_3$~\cite{Yanase_UPt3_Weyl}. 
The neighboring Cooper pairs on ${\bm r}_i$ bonds give equivalent amplitude of $d$-wave and $f$-wave 
components in Eqs.~(\ref{Gamma_1}) and (\ref{Gamma_2}). 
The $d$-wave order parameter plays a particularly important role on the superconducting gap at the BZ face, $k_z =\pi$. 
Later we show that the TNSC is induced by the $d$-wave component.

\begin{figure}[htbp]
\begin{center}
\includegraphics[width=60mm]{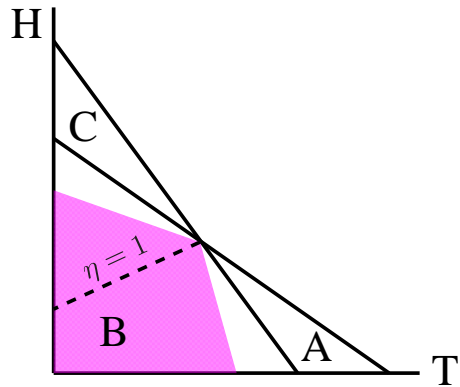}
\caption{(Color online)
Multiple superconducting phases of UPt$_3$ in the magnetic field-temperature plane~\cite{Sauls,Joynt}. 
The A-phase is identified as a TNSC. 
The shaded region shows the Weyl superconducting phase~\cite{Yanase_UPt3_Weyl}. 
Pair creation of Weyl nodes occurs at the phase boundary. 
The dashed line indicates a topological phase transition, which is discussed in Sec.~\ref{sec:magnetic-rotation}. 
} 
\label{phasediagram}
\end{center}
\end{figure}

Now we review the multiple superconducting phases in UPt$_3$. 
Three thermodynamically distinguished superconducting phases are illustrated 
in Fig.~\ref{phasediagram}~\cite{Fisher,Bruls,Adenwalla,Sauls,Joynt}. 
The A-, B-, and C-phases are characterized by the ratio of two-component order parameters 
$\eta=\eta_2/i\eta_1$ summarized in Table.~I. 
A pure imaginary ratio of $\eta_1$ and $\eta_2$ in the B-phase implies the chiral superconducting state which 
maximally gains the condensation energy. Owing to the $p$-wave component, the B-phase is non-unitary. 
It has been considered that the A- and C-phases are stabilized by weak symmetry breaking of hexagonal structure, 
possibly induced by weak antiferromagnetic order~\cite{Sauls,Joynt,Aeppli,Hayden}. 
We here assume that the A-phase is the $\Gamma_2$ state ($\eta=\infty$), 
while the C-phase is the $\Gamma_1$ state ($\eta=0$), and assume non-negative $\eta \ge 0$ without loss of generality. 

\begin{table}[htbp]
{\renewcommand\arraystretch{1.2}
  \begin{tabular}{c|c}
% & $\eta$ 
%\\ 
\hline
A-phase & $|\eta|=\infty$ 
\\ \hline
B-phase & $0 \le |\eta| \le \infty $ 
\\ \hline
C-phase & $\eta=0$ 
\\ 
 \hline
  \end{tabular}
}
  \caption{Range of the parameter $\eta$ in the A-, B-, and C-phases of UPt$_3$~\cite{Joynt,Sauls}. 
}
  \label{tab1}
\end{table}

Contrary to the experimental indications for the $E_{\rm 2u}$-pairing state mentioned before, 
a recent thermal conductivity measurement~\cite{Machida-Izawa} has been interpreted in terms of the 
$E_{\rm 1u}$ symmetry of the {\it orbital part} of order parameter. 
However, this interpretation is not incompatible with the $E_{\rm 2u}$ symmetry of total order parameter. 
For instance, the basis functions, Eqs.~\eqref{Gamma_1} and \eqref{Gamma_2}, include components 
$p_x({\bm k})s_x - p_y({\bm k})s_y$ and $p_y({\bm k})s_x + p_x({\bm k})s_y$, where the orbital part $p_x({\bm k})$ and $p_y({\bm k})$ 
belong to the $E_{\rm 1u}$ symmetry. 
Although in Ref.~\onlinecite{Machida-Izawa} the superconducting state with TRS has been discussed 
along with a theoretical proposal~\cite{Tsutsumi2012}, the spin part of order parameter can not 
be deduced from thermal conductivity measurements. Thus, we here assume the $E_{\rm 2u}$-pairing state.

For clarity of discussions for topological properties, 
we carry out the unitary transformation for the BdG Hamiltonian. 
When the model Eq.~(\ref{eq:model}) is represented in the Nambu space 
\begin{align}
{\cal H}_{\rm BdG} = \frac{1}{2} \sum_{{\bm k}} \hat{c}_{\bm k}^\dagger \hat{H}_{\rm BdG}({\bm k}) \hat{c}_{\bm k},
\end{align}
with 
\begin{align}
\hat{c}_{\bm k} = \left(c_{{\bm k}1\uparrow},c_{{\bm k}2\uparrow},c_{{\bm k}1\downarrow},c_{{\bm k}2\downarrow},c_{-{\bm k}1\uparrow}^{\dag},c_{-{\bm k}2\uparrow}^\dag,c_{-{\bm k}1\downarrow}^\dag,c_{-{\bm k}2\downarrow}^\dag, \right)^{\rm T},
\end{align}
the BdG matrix $\hat{H}_{\rm BdG}({\bm k})$ in this form does not satisfy the periodicity compatible with  
the first BZ. 
To avoid this difficulty, we represent the BdG Hamiltonian by 
\begin{align}
\tilde{H}_{\rm BdG}({\bm k}) =
U({\bm k}) \hat{H}_{\rm BdG}({\bm k}) U({\bm k})^{\dag}, 
\label{unitary_transformation}
\end{align}
using the unitary matrix 
\begin{align}
U({\bm k}) = 
\left(
\begin{array}{cc}
1 & 0 \\
0 & e^{i{\bm k} \cdot {\bm \tau}} \\
\end{array}
\right)_{\sigma}
\otimes
s_0
\otimes
\tau_0. 
\end{align}
By choosing the translation vector, ${\bm \tau}=(0,-\frac{1}{\sqrt{3}},\frac{1}{2})$, 
$\tilde{H}_{\rm BdG}({\bm k})$ is periodic with respect to the translation ${\bm k} \rightarrow {\bm k} + {\bm K}$ 
for any reciprocal lattice vector ${\bm K}$. 
The transformed BdG Hamiltonian has the same form as Eq.~(\ref{eq:model}), although the inter-sublattice 
components acquire the phase factor  
\begin{align}
& a(\k) \rightarrow \tilde{a}(\k) \equiv a(\k) e^{- i{\bm k} \cdot {\bm \tau}} , 
\\
& f_i(\k) \rightarrow \tilde{f}_i(\k) \equiv f_i(\k) e^{- i{\bm k} \cdot {\bm \tau}} , 
\\
& d_i(\k) \rightarrow \tilde{d}_i(\k) \equiv d_i(\k) e^{- i{\bm k} \cdot {\bm \tau}}. 
\end{align}

\section{Topological surface states}

\begin{figure*}[htbp]
\begin{center}
\includegraphics[width=180mm]{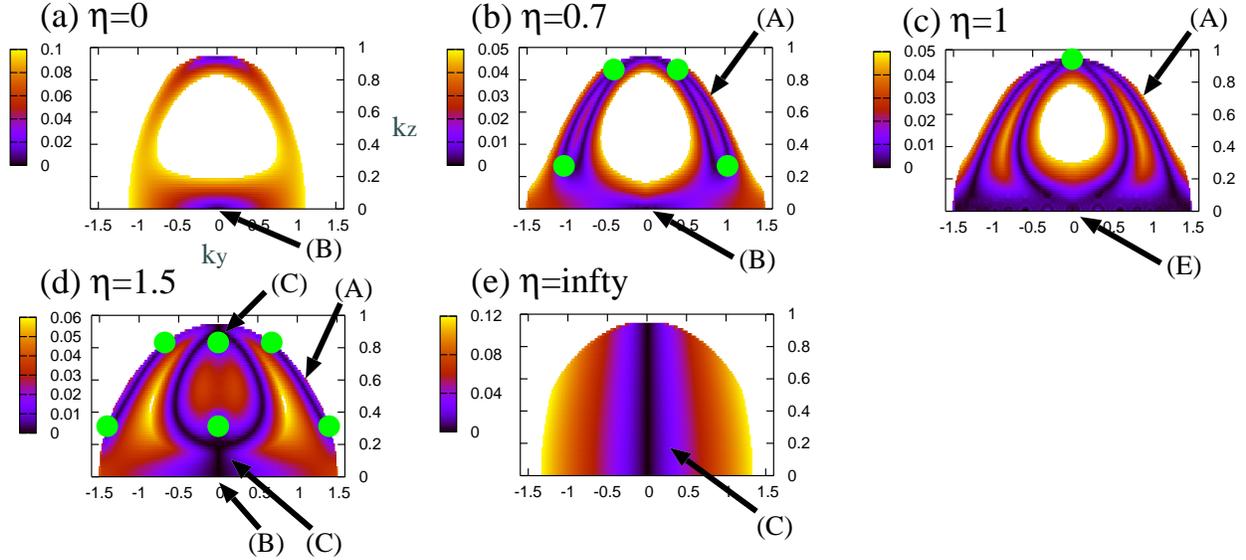}
\caption{(Color online) Energy of surface states on the (100)-surface. 
We impose open boundary condition along the [100]-direction and periodic boundary condition along the other directions. 
The lowest excitation energy of BdG quasiparticles [$\equiv {\rm min} |E(\k_{\rm sf})|$] 
as a function of the surface momentum ${\bm k}_{\rm sf} = (k_y, k_z)$ is shown. 
Parameters $(t,t_z,t',\alpha,\mu,\Delta,\delta)=(1,4,1,0,16,4,0.02)$ are assumed so that 
the $\Gamma$-FS is reproduced. 
(a) C-phase ($\eta=0$), (b)-(d) B-phase ($\eta=0.7$, $1$, and $1.5$), and (e) A-phase ($\eta=\infty$). 
Arrows with characters (A), (B), (C) and (E) indicate surface states clarified  
in Secs.~\ref{sec:Weyl}, \ref{sec:mirror}, \ref{sec:glide-AIII}, and \ref{sec:magnetic-rotation}, respectively. 
The green circles show the projections of Weyl point nodes. 
} 
\label{eGFSedgestate}
\end{center}
\end{figure*}

\begin{figure*}[htbp]
\begin{center}
\includegraphics[width=180mm]{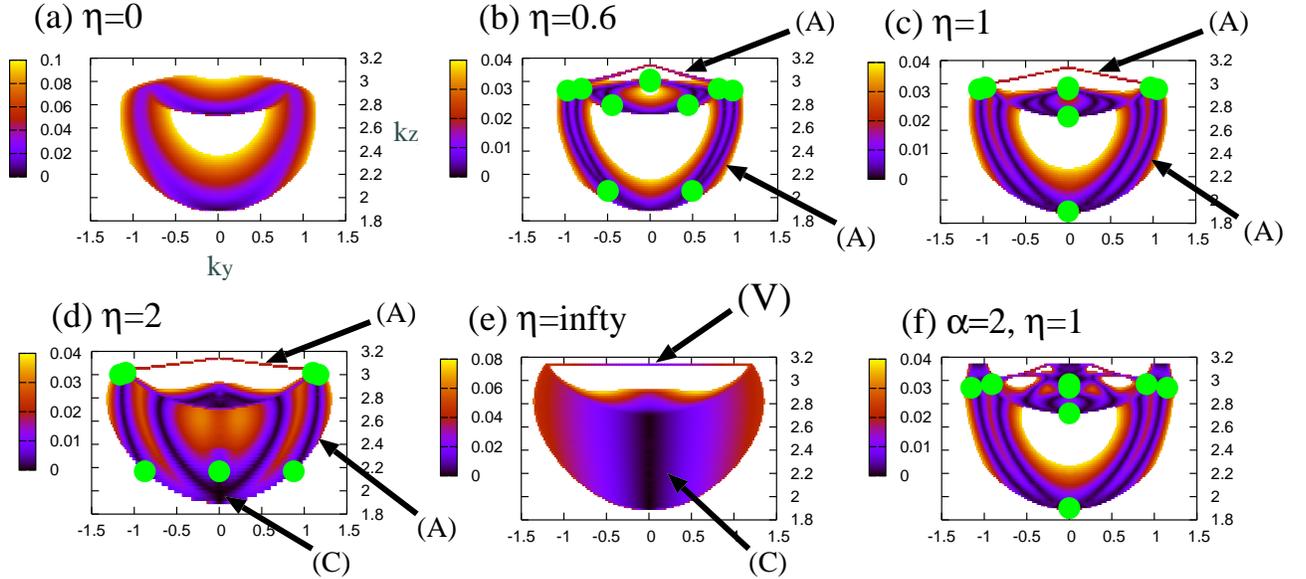}
\caption{(Color online) (a)-(e) Energy of surface states on the (100)-surface 
for parameters reproducing the paired $A$-FSs, $(t,t_z,t',\alpha,\mu,\Delta,\delta)=(1,-4,1,0,12,0.7,0.04)$. 
(a) C-phase ($\eta=0$), (b)-(d) B-phase ($\eta=0.6$, $1$, and $2$), and (e) A-phase ($\eta=\infty$). 
We choose $\alpha=2$ in (f) while the other parameters are the same as (c). 
Comparison between (c) and (f) reveals the effect of the SOC. 
Arrows with characters (V), (A), and (C) indicate surface states discussed 
in Secs.~\ref{sec:glide-DIII}, \ref{sec:Weyl}, and \ref{sec:glide-AIII}, respectively. 
The green circles show the projections of Weyl point nodes. 
} 
\label{ZFSedgestate}
\end{center}
\end{figure*}

We calculate the energy spectrum of quasiparticles with surface normal to the (100)-axis, $E(\k_{\rm sf})=E(k_y,k_z)$, 
because the nonsymmorphic glide symmetry is preserved there. Both glide and screw symmetry 
are broken in the other surface directions.
Figures~\ref{eGFSedgestate} and \ref{ZFSedgestate} show results for the $\Gamma$-FS and $A$-FSs, respectively. 
The black regions represent the zero energy surface states. 
It is revealed that a variety of zero energy surface states appear on the (100)-surface in the A-, B-, and C-phases. 
We clarify the topological protection of these surface states below. 
Indeed, all of the zero energy surface states are topologically protected. 
In Figs.~\ref{eGFSedgestate} and \ref{ZFSedgestate}, the topological surface states discussed in 
Secs.~\ref{sec:glide-DIII} and \ref{sec:Weyl} - \ref{sec:magnetic-rotation} are labeled by 
(\ref{sec:glide-DIII}) and (A)-(E), respectively.

The most panels of Figs.~\ref{eGFSedgestate} and \ref{ZFSedgestate} show the results for $\alpha=0$ 
by neglecting the SOC. Most of the surface states are indeed robust against the SOC. 
Exceptionally, the surface states around $k_{z}=\pi$ are affected by the SOC, 
because the nodal bulk excitations may be induced by the SOC~\cite{Yanase_UPt3_Weyl,Kobayashi-Yanase-Sato}. 
For our choice of parameters, the bulk excitation gap remains finite at $k_z=\pi$ for $\alpha=1$ although the gap 
may be suppressed for $\alpha=2$. 
Thus, we show the surface states for $\alpha=2$ in Fig.~\ref{ZFSedgestate}(f) for a comparison.   
The gapless bulk excitations which are not shown for $\alpha=0$ [Fig.~\ref{ZFSedgestate}(c)] are observed around the surface BZ boundary.

One of the main results of this paper is a signature of TNSC in UPt$_3$, 
that is indicated by the label (V) in Fig.~\ref{ZFSedgestate}(e). 
This surface state is robust against the SOC unless the bulk excitation gap is closed. 
According to the first principles band structure calculation, the band splitting by the SOC is tiny 
along the $A$-$H$ lines~\cite{Joynt} and significantly decreased by the mass renormalization factor~\cite{Maruyama-Yanase2015}, 
$z \sim 1/100$ in UPt$_3$~\cite{Joynt}.  
Thus, it is reasonable to assume a small SOC leading to the gapped bulk excitations at the BZ face.  
This assumption is compatible with the recent field-angle-dependent thermal conductivity measurement 
which has shown nodal lines/points lying away from the BZ face~\cite{Machida-Izawa}.

In the next section, superconducting phases of 3D DIII class with additional glide symmetry 
are classified on the basis of the $K$-theory, and the topological invariants are derived. 
In Sec.~\ref{sec:glide-DIII}, we show that a surface state labeled by (V) is protected 
by the strong topological index characterizing the TNSC. 
The topological protection of other surface states is revealed in Sec.~\ref{sec:low-dimension}.

\section{Classification of class DIII superconductors with glide symmetry}
\label{sec:classification}

Topological classification of TNSC is carried out for both glide-even and glide-odd superconducting states 
of DIII class. For simplicity, the cubic first BZ with volume $(2\pi)^3$ is assumed in this section. 
We do not rely on any specific model, and therefore, the results obtained in this section 
are valid for all the superconducting states preserving the glide symmetry and TRS.

\subsection{Glide-even superconductor}\label{sec:glide-even}

First, we study glide-even superconducting states. 
The $\Gamma_2$-state (A-phase) of UPt$_3$ corresponds to this case. 
The symmetries for the BdG Hamiltonian are summarized as 
\begin{align}
&C {\cal H}(\bk) C^{-1} = -{\cal H}(-\bk), &&  C = \tau_x K, 
\label{glide-even-algebra1} \\
&T {\cal H}(\bk) T^{-1} = {\cal H}(-\bk), &&  T = i s_y K, 
\label{glide-even-algebra2} \\
&G(\bk) {\cal H}(\bk) G^{-1}(\bk) = {\cal H}(m_y\bk), && G(m_y \bk) G(\bk) = - e^{-i k_z}, 
\label{glide-even-algebra3} \\
&T G(\bk) = G(-\bk) T, &&  C G(\bk) = G(-\bk) C,  
\label{glide-even-algebra4} 
\end{align}
where $m_y \bk = (k_x, -k_y, k_z)$ is the momentum flipped by glide operation, 
and $K$ is the complex conjugate. 
The stable classification of bulk superconductors is given by the $K$-theory 
over the bulk 3D BZ torus with symmetries (\ref{glide-even-algebra1})-(\ref{glide-even-algebra4}). 
From Ref.~\onlinecite{Shiozaki2016}, the result is 
\begin{equation}\begin{split}
&({\rm The\ stable\ classification\ of\ bulk\ gapped\ SCs}) \\
&= \underbrace{\bf Z_2}_{(k_x,k_y,k_z)} \oplus \underbrace{\Z_2 \oplus {\bf Z_2}}_{(k_x,k_z)} \oplus \underbrace{\bf Z_2}_{(k_y,k_z)} \oplus \underbrace{\Z_2}_{(k_z)}. 
\end{split}\label{eq:diii+gr_even_bulk}\end{equation}
The bold style ${\bf Z_2}$ expresses an emergent topological phase which disappears if the glide symmetry is broken.
Each underbrace represents the momentum dependence of generating Hamiltonian. 
For instance, $\underbrace{\Z_2}_{(k_x,k_z)}$ means that the generating Hamiltonian of the $\Z_2$ phase can be $k_y$-independent, 
that is, the stacking of layered Hamiltonians $H_y(k_x,k_z)$ in the $xz$-plane along the $y$-direction. 
We focus on the gapless states on the surfaces preserving the glide symmetry, i.e.\ $x=$ constant surface. 
The classification of the surface gapless states is given by a similar $K$-theory over the surface 2D BZ torus 
under the same symmetries (\ref{glide-even-algebra1})-(\ref{glide-even-algebra4}) with the $k_x$-direction excluded. 
The bulk-boundary correspondence holds:~\cite{Shiozaki-Sato-Gomi2017} the $K$-group for the surface gapless states is given by the 
direct summand of which generators are dependent on $k_x$ in Eq.\ (\ref{eq:diii+gr_even_bulk}). 
Thus, 
\begin{align}
& \left(\begin{array}{ll}
{\rm The\ classification\ of\ gapless\ states} \\
{\rm on\ the\ } x= {\rm constant\ surface} 
\end{array}\right)
\nonumber \\
& = \underbrace{\bf Z_2}_{(k_x,k_y,k_z)} \oplus \underbrace{\Z_2 \oplus {\bf Z_2}}_{(k_x,k_z)}. 
\end{align}
All the three $\Z_2$ invariants relevant to the surface gapless states are constructed on the $k_z=\pi$ plane~\cite{Shiozaki2016}.
At $k_z = \pi$, the glide symmetry is reduced into the mirror symmetry 
\begin{align}
&G(k_x,k_y,\pi) {\cal H}(k_x,k_y,\pi) G^{-1}(k_x,k_y,\pi) = {\cal H}(k_x,-k_y,\pi), 
\\ 
& G(k_x,-k_y,\pi) G(k_x,k_y,\pi) = 1, 
\\
&T G(k_x,k_y,\pi) = G(-k_x,-k_y,\pi) T, 
\\
& C G(k_x,k_y,\pi) = G(-k_x,-k_y,\pi) C. 
\end{align}
On the $k_y = \Gamma_y \equiv 0, \pi$ lines, since the TRS and the particle-hole symmetry (PHS) commute with the glide symmetry, 
we can define $\Z_2$ invariant $\nu(\Gamma_y,\pm) \in \{0,1\}$ of one-dimensional (1D) class DIII SCs 
for each glide-subsectors $G(k_x,\Gamma_y,\pi) = \pm 1$, 
\begin{widetext}
\begin{align*}
\nu(\Gamma_y,\pm) 
= \frac{i}{\pi} \oint_0^{2 \pi} d k_x \sum_n \Braket{u^{(I)}_{\pm,n}(k_x,\Gamma_y,\pi) | \partial_{k_x} | u^{(I)}_{\pm,n}(k_x,\Gamma_y,\pi)} \ \ ({\rm mod\ }2), && 
(\Gamma_y = 0, \pi), 
\end{align*}
\end{widetext}
where $u^{(I)}_{\pm,n}(k_x,\Gamma_y,\pi)$ represents one of the Kramers pair of occupied states in the glide-subsector 
$G(k_x,\Gamma_y,\pi) = \pm 1$. 
Noticing that the combined symmetries $TG(k_x,k_y,\pi)$ and $CG(k_x,k_y,\pi)$, 
\begin{align}
&\left[TG(k_x,k_y,\pi)\right] {\cal H}(k_x,k_y,\pi) \left[TG(k_x,k_y,\pi)\right]^{-1} \nonumber \\ &= {\cal H}(-k_x,k_y,\pi),
\\ 
& TG(-k_x,k_y,\pi)TG(k_x,k_y,\pi) = -1, 
\\
&\left[CG(k_x,k_y,\pi)\right] {\cal H}(k_x,k_y,\pi) \left[CG(k_x,k_y,\pi)\right]^{-1} \nonumber \\ &= -{\cal H}(-k_x,k_y,\pi), 
\\
& CG(-k_x,k_y,\pi)CG(k_x,k_y,\pi) = 1, 
%&(TG(\pi,k_y,-k_z)) (CG(\pi,k_y,k_z)) = (CG(\pi,k_y,-k_z)) (TG(\pi,k_y,k_z)), 
\end{align}
indicate the emergent class DIII symmetry for all $k_y$, 
we have a constraint 
\begin{align}
\nu(0,+) + \nu(0,-) = \nu(\pi,+) + \nu(\pi,-), \ \ ({\rm mod\ }2). 
\end{align}
Because of this emergent class DIII symmetry, all the surface states on the $k_z= \pi$ plane show two-fold degeneracy. 
The three kinds of surface states may be generated by 
\begin{widetext}
\begin{align}
&\underbrace{\bf Z_2}_{(k_x,k_y,k_z)} : && \bigl(\nu(0,+), \nu(0,-); \nu(\pi,+), \nu(\pi,-)\bigr) = (1,1;0,0) \ {\rm or\ } (0,0;1,1), 
\label{glide_strong_Z2}
\\
&\underbrace{\Z_2}_{(k_x,k_z)} : && \bigl(\nu(0,+), \nu(0,-); \nu(\pi,+), \nu(\pi,-)\bigr) = (1,0;1,0), (1,0;0,1), (0,1;1,0) \ {\rm or\ } (0,1;0,1), 
\label{glide_weak_Z2flat}
\\
&\underbrace{\bf Z_2}_{(k_x,k_z)} : && \bigl(\nu(0,+), \nu(0,-); \nu(\pi,+), \nu(\pi,-)\bigr) = (1,1;1,1). 
\label{glide_weak_Z2}
\end{align}
\end{widetext}
Here, $\underbrace{\Z_2}_{(k_x,k_z)}$ shows the flat surface band on the $k_z = \pi$ plane. 
The $\underbrace{\bf Z_2}_{(k_x,k_y,k_z)}$ is the strong index of TNSC, which is denoted as 
glide-$\Z_2$ invariant, $\nu_{\rm G} \equiv \underbrace{\bf Z_2}_{(k_x,k_y,k_z)}$. 
It is given by  
\begin{align}
\nu_{\rm G} = \nu(0,+) \nu(0,-) - \nu(\pi,+) \nu(\pi,-) \,\,\,\,\, ({\rm mod} \,\,\, 2). 
\label{glide_Z2}
\end{align}
Later, we show that the glide-$\Z_2$ invariant $\nu_{\rm G}$ is nontrivial in the A-phase of UPt$_3$.

\subsection{Glide-odd superconductor}\label{glide-odd}

Next, we study glide-odd superconducting states, which may be realized in the $\Gamma_1$-state of UPt$_3$ (C-phase). 
Symmetries for the BdG Hamiltonian are 
\begin{align}
&T {\cal H}(\bk) T^{-1} = {\cal H}(-\bk), && T = i s_y K, 
\label{glide-odd-algebra1}
\\
&C {\cal H}(\bk) C^{-1} = -{\cal H}(-\bk), && C = \tau_x K, 
\label{glide-odd-algebra2}
\\
&G(\bk) {\cal H}(\bk) G^{-1}(\bk) = {\cal H}(m_y\bk), && G(m_y \bk) G(\bk) = -e^{-i k_z}, 
\label{glide-odd-algebra3}
\\
&T G(\bk) =G(-\bk) T, && C G(\bk) = -G(-\bk) C. 
\label{glide-odd-algebra4}
\end{align}
From Ref.~\onlinecite{Shiozaki2016}, the $K$-theory classification of the bulk reads 
\begin{equation}\begin{split}
&({\rm The\ stable\ classification\ of\ bulk\ gapped\ SCs}) \\
&= \underbrace{\Z \oplus {\bf Z_2}}_{(k_x,k_y,k_z)} \oplus \underbrace{\bf Z_4}_{(k_x,k_z)} \oplus \underbrace{\Z_2 \oplus {\bf Z_2}}_{(k_y,k_z)} \oplus \underbrace{\Z_2}_{(k_z)}. 
\end{split}\label{eq:diii+gr_odd_bulk}\end{equation}
The bold-style indices express emergent topological phases which requires the glide symmetry. 
From the bulk-boundary correspondence, 
it holds that
\begin{align}
& \left(\begin{array}{ll}
{\rm The\ classification\ of\ surface\ states} \\
{\rm on\ the\ } x= {\rm constant\ surface} 
\end{array}\right)
%& ({\rm The\ classification\ of\ surface\ states\ \\ & on\ the\ } x= {\rm constant\ surface} ) 
\nonumber \\ &
= \underbrace{\Z \oplus {\bf Z_2}}_{(k_x,k_y,k_z)} \oplus \underbrace{\bf Z_4}_{(k_x,k_z)}. 
\end{align}

The 3D $\underbrace{\Z}_{(k_x,k_y,k_z)}$ index is the ordinary winding number~\cite{Schnyder}, 
\begin{align}
N := \frac{1}{48 \pi^2} \int {\rm tr} \Gamma ({\cal H}^{-1} d {\cal H})^3, && \Gamma = i T C. 
\end{align}
By imposing the glide symmetry, we have two $\Z_4$ invariants $\theta(\Gamma_y = 0, \pi) \in \{0,1,2,3\}$ 
on the glide invariant $k_y = 0$ and $\pi$ planes~\cite{Shiozaki2016}, 
\begin{widetext}
\begin{align}
\theta(\Gamma_y):= \frac{2 i}{\pi} 
\Big[ \oint_0^{2 \pi} d k_x {\rm tr} {\cal A}^{(I)}_{+}(k_x, \Gamma_y, \pi) + \frac{1}{2} \int_0^{\pi} d k_z \oint_0^{2 \pi} d k_x {\rm tr} {\cal F}_+(k_x,\Gamma_y,k_z) \Big] \ \ ({\rm mod\ } 4), && 
(\Gamma_y = 0, \pi). 
\end{align}
\end{widetext}
Here, ${\cal A}^{(I)}_{+}(k_x, \Gamma_y, \pi)$ and ${\cal F}_+(k_x,\Gamma_y,k_z)$ are the Berry connection of one of Kramers pair of occupied states and 
the Berry curvature of the occupied states, respectively, with the positive glide eigenvalue $G(k_x,\Gamma_y,\pi) = 1$. 
In modulo 2, $\theta(\Gamma_y)$ is recast into the $\Z_2$ invariant at $(k_x,\Gamma_y,k_z=0)$ lines as~\cite{Shiozaki2016}
\begin{align}
\theta(\Gamma_y):= \frac{i}{\pi} \oint_0^{2 \pi} d k_x {\rm tr} {\cal A}_{+}(k_x, \Gamma_y, 0) \ \ ({\rm mod\ } 2), 
%&& \Gamma_y = 0, \pi. 
\end{align}
by the Stokes' theorem. 

The three invariants $\{ N, \theta(0), \theta(\pi) \}$ are not independent, since there is a constraint 
\begin{align}
N + \theta(0) + \theta(\pi) = 0 \ \ ({\rm mod\ }2), 
\label{eq:constraint}
\end{align}
which can be understood as follows. 
On the $k_z=0$ plane, the $\Z_2$ invariant $\nu = \theta(0) + \theta(\pi) \ ({\rm mod\ }2)$ is equivalent 
to the 2D class DIII $\Z_2$ invariant. 
Since we can show that the existence of odd numbers of Majorana cones is allowed only on the $k_z=0$ plane, 
$N$ (mod 2) is also equivalent to the $\Z_2$ invariant. 
Therefore, $N = \nu \ ({\rm mod  \  2})$, which implies Eq.~(\ref{eq:constraint}).

\section{Topological nonsymmorphic superconductivity in A-phase}\label{sec:glide-DIII}

\begin{figure}[htbp]
\begin{center}
\includegraphics[width=85mm]{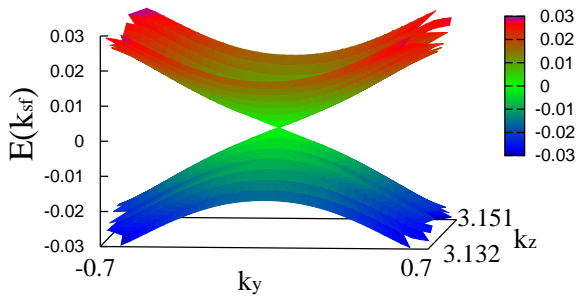}
\caption{(Color online) 
Double Majorana cone in the A-phase. 
Energy spectrum of (100)-surface states around $\k_{\rm sf}=(0,\pi)$ is shown. 
Parameters are the same as Fig.~\ref{ZFSedgestate}(e) for the paired $A$-FSs. 
} 
\label{glide-cone}
\end{center}
\end{figure}

Now we go back to the superconductivity in UPt$_3$. 
Let's focus on the surface zero mode at $\bk_{\rm sf} =(0,\pi)$ in the TRS invariant A-phase. 
Naturally, the $A$-FSs are considered in this section. 
The surface states labeled by (V) in Fig.~\ref{ZFSedgestate}(e) have the spectrum shown in Fig.~\ref{glide-cone}. 
As we proved in Sec.~\ref{sec:glide-even}, the quasiparticle states are two-fold degenerate on the 
$\k_{\rm sf} = (k_y, \pi)$ line in the glide-even A-phase. 
Therefore, the spectrum of surface states shows double Majorana cone with four zero energy states at $\bk_{\rm sf} =(0,\pi)$.
In this section we show that the surface double Majorana cone is protected by the strong $\mathbb{Z}_2$ index $\nu_G$
for glide-even TNSCs, which has been introduced in Eq.~\eqref{glide_Z2}.

The glide symmetry of $P6_3/mmc$ space group is $G^{xz} = \{M^{xz}|\frac{z}{2}\}$ composed of mirror reflection 
and half translation along the $z$-axis. Thus, the nonsymmorphic glide operator is intrinsically $k_z$-dependent. 
We have an operator for the normal part Hamiltonian, $G^{xz}(k_z) = i s_y \sigma_x V_{\sigma}(k_z)$, where 
\begin{align}
&
V_{\sigma}(k_z) = 
%s_0 \otimes
\left(
\begin{array}{cc}
1 & 0 \\
0 & e^{-i k_z} \\
\end{array}
\right)_{\sigma}, 
%\otimes \tau_0, 
\end{align}
acts in the sublattice space. 
The superconducting state preserves the glide symmetry in the TRS invariant A- and C-phases, although 
the glide symmetry is spontaneously broken in the B-phase. 
The glide operator in the Nambu space depends on the glide-parity of the superconducting state; 
$G^{xz}_{\rm BdG}(k_z) = G^{xz}(k_z) \tau_0$ in the glide-even A-phase while 
$G^{xz}_{\rm BdG}(k_z) = G^{xz}(k_z) \tau_z$ in the glide-odd C-phase. 
Then, the BdG Hamiltonian respects the glide symmetry 
\begin{align}
& %\hspace{-7mm}
G^{xz}_{\rm BdG}(k_z) \tilde{H}_{\rm BdG}(\k) G^{xz}_{\rm BdG}(k_z)^{-1} = \tilde{H}_{\rm BdG}(k_x,-k_y,k_z).
\nonumber \\
\label{glide-symmetry}
\end{align}
The symmetries satisfy the algebra (\ref{glide-even-algebra1})-(\ref{glide-even-algebra4}) 
and (\ref{glide-odd-algebra1})-(\ref{glide-odd-algebra4}) in the A-phase and C-phase, respectively.

\subsection{Glide-$\Z_2$ invariant}\label{1D_invariant}

\begin{figure}[htbp]
\begin{center}
\includegraphics[width=70mm]{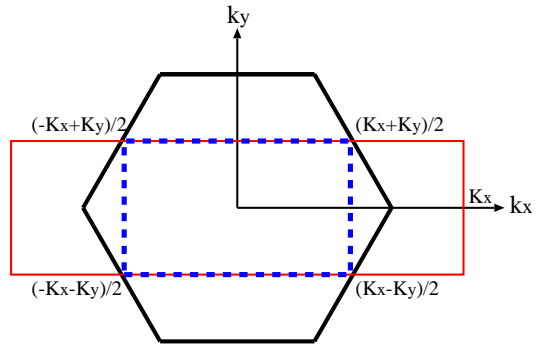}
\caption{(Color online) Unfolded BZ (solid line) and folded BZ (dashed line) 
projected onto a $k_z =$ constant plane. 
The latter is compatible with the surface BZ. 
$\K_x = 2 \pi \hat{x}$ and $\K_y = \frac{2 \pi}{\sqrt{3}} \hat{y}$ are reciprocal lattice vectors of 
the folded BZ. 
} 
\label{folded_BZ}
\end{center}
\end{figure}

As we showed in Sec.~\ref{sec:glide-even}, only the 2D plane at $k_z=\pi$ determines the topological properties of 
the glide-even A-phase. From Eq.~(\ref{glide_Z2}), the glide-$\Z_2$ invariant $\nu_G$ is given by 
the 1D $\Z_2$ invariant of DIII class, $\nu(\Gamma_y,\pm)$. 
When we choose the rectangular BZ shown in Fig.~\ref{folded_BZ}, we have $\Gamma_y=0,\pi/\sqrt{3}$. 
For our choice of parameters, the FSs do not cross a line $k_y=\pi/\sqrt{3}$ on the BZ face. 
Therefore, $\nu(\pi/\sqrt{3},\pm)$ is trivial, and the glide-$\Z_2$ invariant is obtained by evaluating $\nu(0,\pm)$. 
Below, we show that $\nu(0,\pm) = 1$, and thus, the glide-$\Z_2$ invariant is nontrivial.

First, the Hamiltonian is block-diagonalized by using the basis diagonal for $G^{xz}_{\rm BdG}(\pi) = s_y \sigma_y \tau_0$, 
\begin{align}
\tilde{H}_{\rm BdG}(k_x,0,\pi) = \tilde{H}_{1}^{\rm 1d}(k_x) \oplus \tilde{H}_{-1}^{\rm 1d}(k_x),
\end{align} 
on the 1D BZ $k_x \in [-2 \pi,2 \pi]$. 
%and thus $\tilde{H}_{\rm BdG}(k_x,0,\pi) = \tilde{H}_{1}^{\rm 1d}(k_x) \oplus \tilde{H}_{-1}^{\rm 1d}(k_x)$. 
%
The glide-subsector with eigenvalue $\lambda_{\rm G}=\pm 1$ is obtained as, 
\begin{align}
\tilde{H}_{\pm 1}^{\rm 1d}(k_x)  &=
\left(
\begin{array}{cc}
\hat{H}_{\pm 1}^{(0)}(k_x)  & \hat{\Delta}(k_x) \\
\hat{\Delta}(k_x)^\dag  &  - \hat{H}_{\pm 1}^{(0)}(-k_x)^T \\
\end{array}
\right), 
\end{align}
with
\begin{align}
\hat{H}_{\pm 1}^{(0)}(k_x)  &=
\left(
\begin{array}{cc}
\xi^{\rm 1d}(k_x)  & \mp \alpha {\bm g}^{\rm 1d}(k_x)  \\
\mp \alpha {\bm g}^{\rm 1d}(k_x) & \xi^{\rm 1d}(k_x) \\
\end{array}
\right),
\\
\hat{\Delta}(k_x)  &= i
\Delta
\left(
\begin{array}{cc}
i \delta p_{x}^{\rm 1d}(k_x)  & - d_{xz}^{\rm 1d}(k_x)  \\
- d_{xz}^{\rm 1d}(k_x) & i \delta p_{x}^{\rm 1d}(k_x) \\
\end{array}
\right). 
\end{align}
% 
%where $\xi^{\rm 1d}(k_x)=\xi(k_x,0,\pi)$, ${\bm g}^{\rm 1d}(k_x) = {\bm g}(k_x,0,\pi)$, 
We defined 
$p_{x}^{\rm 1d}(k_x) \equiv p_x(k_x,0,\pi) = \sin k_x + \sin \frac{k_x}{2}$ 
and $d_{xz}^{\rm 1d}(k_x) \equiv d_{xz}(k_x,0,\pi) = -\sqrt{3} \sin \frac{k_x}{2}$. 
Thus, the glide-subsector is equivalent to the TRS invariant $p$-wave SC. 
It is easy to confirm that both TRS and PHS are preserved in each glide-subsector as expected from 
Sec.~\ref{sec:glide-even}. 
In the A-phase we adopt time-reversal operator in the Nambu space $T_{\rm BdG} =i T \tau_z$, 
since the gap function is chosen to be pure imaginary. 
Then, the commutation relation $\left[C, T_{\rm BdG} \right] =0$ is satisfied.

Although the inversion symmetry is broken in the glide-subsector by the SOC, 
we can adiabatically eliminate the SOC as $\alpha \rightarrow 0$, unless the SOC is large enough 
to suppress the superconducting gap~\cite{Yanase_UPt3_Weyl,Kobayashi-Yanase-Sato}.  
Then, the glide-subsector is reduced to the odd-parity spin-triplet SC, and the $\mathbb{Z}_2$ invariant 
is obtained by counting the number of Fermi points $N(\lambda_{\rm G})$ (per Kramers pairs) 
between the time-reversal invariant momentum, $k_x=0$ and $2 \pi$~\cite{Sato2010}. 
Since each glide-subsector represents a single band model with $N(\pm 1)=1$, 
the nontrivial $\mathbb{Z}_2$ invariant, $\nu(0,\pm) =1$ (mod 2), is obtained from the formula 
$(-1)^{\nu(0,\pm)} = (-1)^{N(\pm)}$.

Now we conclude that the glide-$\Z_2$ invariant is nontrivial, namely, $\nu_{\rm G}=1$, 
because 
\begin{align}
\Bigl(\nu(0,+), \nu(0,-); \nu(\pi/\sqrt{3},+), \nu(\pi/\sqrt{3},-)\Bigr) = (1,1;0,0). 
\end{align}
This is the strong topological index characterizing the TNSC with even glide-parity.

It should be noticed that the paired FSs and the sublattice-singlet $d$-wave pairing are essential ingredients. 
Both of them are ensured by the nonsymmorphic space group symmetry (see Secs.~\ref{sec:Dirac-node} and \ref{sec:order-parameter}). 
The pseudospin degree of freedom in the glide-subsector corresponds to the pair of FSs. 
Although the $f$-wave component in the order parameter disappears on the glide invariant plane $k_y =0$, 
the $d$-wave component induces the superconducting gap and gives rise to 1D $\Z_2$ nontrivial superconductivity.

The topological surface state protected by the glide-$\Z_2$ invariant should appear as a signature of the TNSC. 
Because the two glide-subsectors discussed above are TRS invariant and $\mathbb{Z}_2$ nontrivial, 
two Majorana states per subsector, namely, four Majorana states in total, appear 
on the glide invariant (100)-surface. Indeed, the double Majorana cone centered at 
$\k_{\rm sf}=(0,\pi)$ (Fig.~\ref{glide-cone}) is the characteristic topological surface states of the glide-even TNSC.

\begin{table}[htbp]
{\renewcommand\arraystretch{1.2}
\begin{tabular}{c|c|c|c|c|c|c}
& $k_z$ & $\left(G_{\rm BdG}^{xz}\right)^2$ & $\eta_T$ & $\eta_C$ & 1D invariant & 2D invariant \\
\hline
C-phase & 0 & -1 & 1 & -1 & $\mathbb{Z}_2$ & $\mathbb{Z}_2 \oplus \mathbb{Z}_2$  \\ \cline{2-7}
($\eta = 0$) & $\pi$ & 1 & 1 & -1 & $0$ & $0$ \\ \hline
A-phase & 0 & -1 & 1 & 1 & $\mathbb{Z}$ & $\mathbb{Z}$ \\ \cline{2-7}
($\eta = \infty$) & $\pi$ & 1 & 1 & 1 & $\mathbb{Z}_2 \oplus \mathbb{Z}_2$  & $\mathbb{Z}_2$ \\ 
\hline
\end{tabular}
} 
\caption{Classification of 1D and 2D BdG Hamiltonian in the TRS invariant A- and C-phases. 
The low-dimensional Hamiltonian on the basal plane $(k_z=0$) and BZ face ($k_z=\pi$) is classified. 
We show $\left(G_{\rm BdG}^{xz}\right)^2$, $\eta_T$, and $\eta_C$. (Anti-)commutation relations with 
time-reversal and particle-hole operators are represented as 
$T_{\rm BdG} \, G_{\rm BdG}^{xz} = \eta_T \, G_{\rm BdG}^{xz} \, T_{\rm BdG}$ and 
$C \, G_{\rm BdG}^{xz} = \eta_C \, G_{\rm BdG}^{xz} \, C$. 
The right two columns show the 1D topological index on the $(k_y, k_z) =(0,0)$ and $(0,\pi)$ lines and 
the 2D topological index on the $k_z=0$ and $\pi$ planes. 
}
\label{table:glide}
\end{table}

For confirmation, we show the topological indices of 1D Hamiltonian along the $\k = (k_x,0,0)$ and $(k_x,0,\pi)$ lines 
and 2D Hamiltonian on the $k_z=0$ and $\pi$ planes in Table~II. 
For these low-dimensional Hamiltonian, the glide operator is momentum independent, 
and therefore, the topological classification can be carried out without taking care of the nonsymmorphic property. 
The (anti-)commutation relations of symmetry operators are summarized in Table~\ref{table:glide}, 
and accordingly the topological indices are obtained on the basis of the periodic table for symmorphic 
topological crystalline insulators and SCs~\cite{Shiozaki2014}.
Indeed, in the A-phase we have $\mathbb{Z}_2 \oplus \mathbb{Z}_2$ index for 1D Hamiltonian 
on the $\k = (k_x, 0, \pi)$ line, which are nothing but $\nu(0,\pm)$. 
The $\mathbb{Z}_2$ index of 2D Hamiltonian on the $k_z=\pi$ plane is equivalent to the glide-$\Z_2$ invariant 
discussed in this section. 
On the other hand, the $k_z=\pi$ plane is trivial in the glide-odd C-phase, consistent with the absence of 
topological surface states in Fig.~\ref{ZFSedgestate}(a).

\subsection{Folded Brillouin zone}\label{2D_invariant}

For the consistency with classification based on the $K$-theory in Sec.~\ref{sec:classification}, 
we need to consider the folded Brillouin zone compatible with the surface BZ. 
To be specific, the translation symmetry along the [010]-axis is partially broken on the (100)-surface. 
The basic translation vectors on the surface are $(y,z)=(\sqrt{3},0)$ and $(0,1)$, 
and the reciprocal lattice vectors are $\K_y = \frac{2}{\sqrt{3}}\pi \hat{y}$ and 
$\K_z = 2\pi \hat{z}$. Thus, the surface first BZ is a rectangle with 
$k_y \in [-\pi/\sqrt{3},\pi/\sqrt{3})$ and $k_z \in [-\pi,\pi)$. 
We have already adopted the bulk BZ compatible with the surface BZ. However, the 
periodicity with respect to $\K_y$ is lost in the BdG Hamiltonian. 
To satisfy the periodicity, we equate $\k$ with $\k + \K_y$, and accordingly, adopt the folded BZ 
in Fig.~\ref{folded_BZ}.

The folded BdG Hamiltonian is transformed  
\begin{align}
\widehat{H}_{\rm BdG}(\k) =
U_{\rm sf}(\k) 
\left(
\begin{array}{cc}
\tilde{H}_{\rm BdG}(\k) & 0 \\
0 &  \tilde{H}_{\rm BdG}(\k + \K_y) \\
\end{array}
\right)_\rho
U_{\rm sf}(\k)^\dag, 
\end{align}
by the unitary matrix 
\begin{align}
U_{\rm sf}(\k) = 
\frac{1}{\sqrt{2}}
\left(
\begin{array}{cc}
1 & 0 \\
0 & e^{i \k {\bm \tau}'} \\
\end{array}
\right)_\rho
\left(
\begin{array}{cc}
1 & 1 \\
1 & -1 \\
\end{array}
\right)_\rho, 
\label{folded_BdG_Hamiltonian}
\end{align}
with ${\bm \tau}'=(\frac{1}{2}, \frac{\sqrt{3}}{2}, 0)$. 
It is easy to check the periodicity of the folded BdG Hamiltonian, 
$\widehat{H}_{\rm BdG}(\k + \K_i) 
%= \widehat{H}_{\rm BdG}(\k + \K_y) = \widehat{H}_{\rm BdG}(\k + \K_z) 
= \widehat{H}_{\rm BdG}(\k)$ with respect to $\K_x = 2\pi \hat{x}$, $\K_y$, and $\K_z$. 
The glide symmetry is recast,  
\begin{align}
& %\hspace{-7mm}
\widehat{G}^{xz}_{\rm BdG}(\k) \, \widehat{H}_{\rm BdG}(\k) \, \widehat{G}^{xz}_{\rm BdG}(\k)^{-1} = \widehat{H}_{\rm BdG}(k_x,-k_y,k_z), 
\label{glide-symmetry-folded}
\end{align}
with using the glide operator for the folded BdG Hamiltonian, 
\begin{align}
& %\hspace{-7mm}
\widehat{G}^{xz}_{\rm BdG}(\k) = 
G^{xz}_{\rm BdG}(k_z) \otimes 
\left(
\begin{array}{cc}
1 & 0 \\
0 & e^{-i \sqrt{3} k_y} \\
\end{array}
\right)_{\rho}. 
\label{glide-operator-folded}
\end{align}
The TRS and PHS are also preserved, 
\begin{align}
& %\hspace{-7mm}
T_{\rm BdG} \, \widehat{H}_{\rm BdG}(\k) \, T_{\rm BdG}^{-1} = \widehat{H}_{\rm BdG}(-\k), 
\\
& C \, \widehat{H}_{\rm BdG}(\k) \, C^{-1} = - \widehat{H}_{\rm BdG}(-\k). 
\label{TC-symmetry-folded}
\end{align}
The symmetry operators satisfy the following relations, 
\begin{align}
& %\hspace{-7mm}
T_{\rm BdG} \, C = C \, T_{\rm BdG}, 
\\
&
\widehat{G}^{xz}_{\rm BdG}(m \k) \, \widehat{G}^{xz}_{\rm BdG}(\k) = - e^{-i k_z},
\\
&
T_{\rm BdG} \, \widehat{G}^{xz}_{\rm BdG}(\k) = \widehat{G}^{xz}_{\rm BdG}(-\k) \, T_{\rm BdG},
\\
&
C \, \widehat{G}^{xz}_{\rm BdG}(\k) = \pm \widehat{G}^{xz}_{\rm BdG}(-\k) \, C. 
\label{symmetry-class-folded}
\end{align}
The sign $\pm$ in Eq.~(\ref{symmetry-class-folded}) should be chosen in the A-phase and C-phase, respectively. 
Thus, the algebra (\ref{glide-even-algebra1})-(\ref{glide-even-algebra4}) and (\ref{glide-odd-algebra1})-(\ref{glide-odd-algebra4}) 
are satisfied.

The 1D $\Z_2$ invariants in the folded BZ are equivalent to those obtained in the unfolded BZ. 
This fact is simply understood by looking at the surface states. 
The energy spectrum is not changed by the unitary transformation, and 
we have obtained odd number of Majorana cone at $\bk_{\rm sf} =(0,\pi)$ per glide-subsector. 
This fact indicates $\nu(0,\pm) =1$ and $\nu(\pi/\sqrt{3},\pm) =0$ (mod 2). Therefore, the glide-$\Z_2$ invariant 
is nontrivial, that is, $\nu_{\rm G}=1$.

\subsection{Deformation to M\"obius surface state}\label{3D_invariant}

\begin{figure}[htbp]
\begin{center}
\includegraphics[width=75mm]{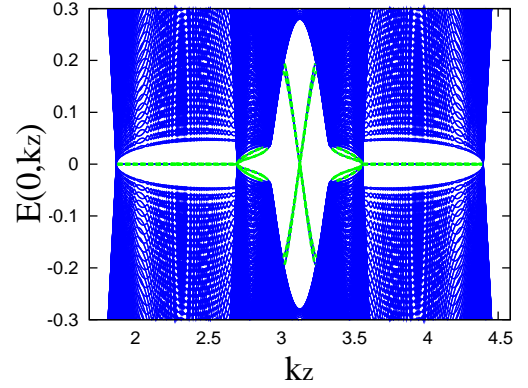}
\caption{(Color online) Energy spectrum on the (100)-surface in the A-phase. Parameters 
$(t,t_z,t',\alpha,\mu,\Delta,\delta)=(1,-4,1,2,12,0.7,0.04)$ reproduce the paired $A$-FSs of UPt$_3$. 
Spectrum on the $\k_{\rm sf}=(0, k_z)$ line is shown. Surface states are highlighted by green lines. 
} 
\label{3DFS_glide}
\end{center}
\end{figure}

The glide-$\Z_2$ invariant $\nu_{\rm G}$ is the strong topological index specifying the gapped TNSC.  
However, the A-phase is actually gapless because of the point nodes of gap function at the poles of 3D FSs. 
Figure~\ref{3DFS_glide} shows the surface spectrum $E(0,k_z)$, and indeed, we observe gapless bulk 
excitations away from the surface BZ boundary $k_z=\pi$ in addition to the double Majorana cone at $k_z=\pi$. 
Therefore, UPt$_3$ A-phase does not realize the characteristic ``M\"obius surface state''~\cite{Shiozaki2015,Shiozaki2016,KHgX,CeNiSn} 
of topological nonsymmorphic insulators/superconductors.

However, the nontrivial glide-$\mathbb{Z}_2$ invariant ensures that the gapped TNSC can be realized 
when the point nodes are removed by some perturbations preserving the symmetry. 
Then, we obtain the M\"obius surface states with keeping the nontrivial glide-$\Z_2$ invariant 
and the associated double Majorana cone. 
In other words, the double Majorana cone around $\k_{\rm sf} = (0,\pi)$ is regarded as a reminiscent of 
the M\"obius surface states of glide-even TNSC. 

\begin{figure}[htbp]
\begin{center}
\includegraphics[width=85mm]{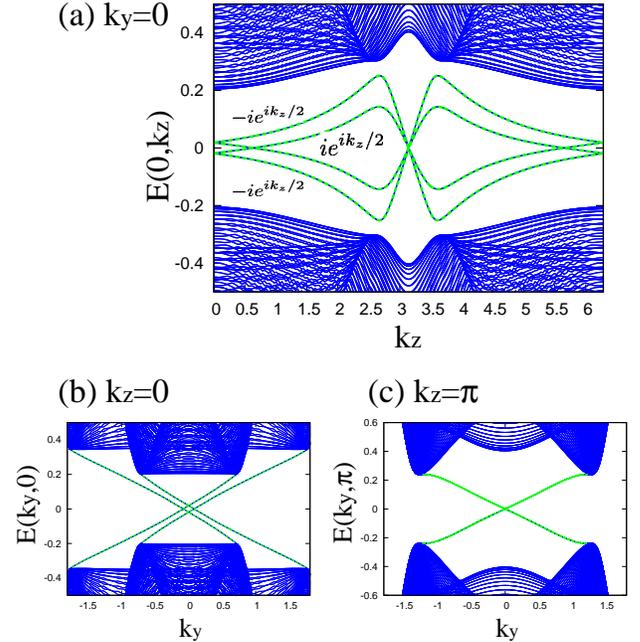}
\caption{(Color online) 
M\"obius surface states in the 3D glide-even TNSC. 
We choose parameters $(t,t_z,t',\alpha,\mu,\Delta,\delta)=(1,0,0.3,0.5,3.5,0.4,0.5)$ for 2D cylindrical FSs. 
Bulk and surface states are shown by blue and green lines, respectively. 
(a) $E(0, k_z)$, (b) $E(k_y, 0)$, and (c) $E(k_y, \pi)$. 
The glide eigenvalues are illustrated in (a). 
} 
\label{2DFS}
\end{center}
\end{figure}

A simple way is to deform the FS to be cylindrical so that the point nodes are removed. 
Then, the surface spectrum in Fig.~\ref{2DFS} is obtained. 
In Fig.~\ref{2DFS}(a), the surface states detached from bulk excitations show the M\"obius 
structure typical of glide-even TNSC. At $\k_{\rm sf}=0$, the Kramers degeneracy is ensured by the TRS. 
The Kramers pair is formed by $\pm i$ glide eigenstates since the TRS and PHS are not preserved 
in the glide-subsector. 
When we look at the $\k_{\rm sf} = (k_y, 0)$ line, Fig.~\ref{2DFS}(b) shows two helical modes protected by the 
mirror Chern number $\nu_{\rm M}^0= 4$, which is introduced in Sec.~\ref{sec:mirror}. 
The nontrivial relationship between the mirror Chern number and the glide-$\mathbb{Z}_2$ invariant 
will be shown elsewhere~\cite{Shiozaki-Yanase2017}.

\subsection{Broken glide symmetry by crystal distortion}\label{sec:broken_glide}

Strictly speaking, the symmetry of crystal structure in UPt$_3$ is still under debate, because 
a tiny crystal distortion has been indicated by a x-ray diffraction measurement~\cite{Walko}. 
The distortion leads to layer dimerization that breaks the glide and screw symmetry. 
Then, the space group is reduced from nonsymmorphic $P6_3/mmc$ to symmorphic $P\bar{3}m1$. 
If the crystal distortion actually occurs in UPt$_3$, the double Majorana cone protected 
by the glide-$\Z_2$ invariant may be gapped.

\begin{figure}[htbp]
\begin{center}
\includegraphics[width=70mm]{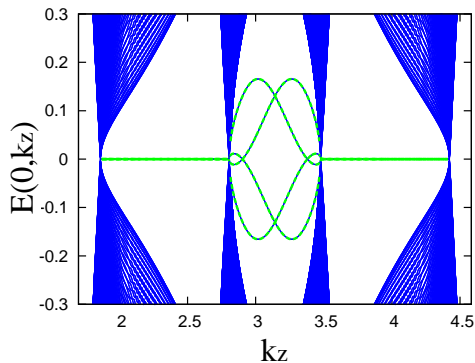}
\caption{(Color online) Energy spectrum on the (100)-surface in the presence of layer dimerization 
that breaks the glide symmetry. Parameters are the same as Fig.~\ref{3DFS_glide}, while the inter-sublattice 
hybridization is replaced by Eq.~(\ref{layer_dimerization}) with $d=0.2$. 
Surface states are highlighted by green lines. 
} 
\label{3DFS_Bglide}
\end{center}
\end{figure}

The layer dimerization makes the inter-sublattice hybridization asymmetric between the $+z$ and $-z$ directions. 
The asymmetry is taken into account by replacing, 
\begin{align}
& a({\bm k}) = 2 t' \cos\frac{k_z}{2} \sum_{i=1,2,3} e^{i{\bm k}_\parallel\cdot{\bm r}_i} 
\nonumber \\
&
\Rightarrow \,\,
t' \left[(1+d) e^{i k_z/2} + (1-d) e^{-i k_z/2} \right] \sum_{i=1,2,3} e^{i{\bm k}_\parallel\cdot{\bm r}_i}.  
\label{layer_dimerization}
\end{align}
The parameter $d$ represents the strength of the layer dimerization. 
For a finite $d$, the double Majorana cone at $\k_{\rm sf}=(0,\pi)$ indeed acquires mass term. 
In Fig.~\ref{3DFS_Bglide}, we show the surface spectrum gapped at $\k_{\rm sf}=(0,\pi)$. 
In the figure, a strong layer dimerization $d=0.2$ is assumed in order to visualize the effect of glide symmetry breaking. 
In reality, the parameter $d$ is expected to be tiny even if it is finite, because the crystal distortion 
reported is small~\cite{Walko}. 
Therefore, the gap in the double Majorana cone may be tiny, and a fingerprint of topological 
glide-$\Z_2$ superconductivity will appear even in the symmorphic $P\bar{3}m1$ structure.

\section{Other topological surface states}\label{sec:low-dimension}

In contrast to toy models, the model specific for the real material shows rich topological properties. 
In Figs.~\ref{eGFSedgestate} and \ref{ZFSedgestate}, we have observed a variety of surface states 
other than the double Majorana cone discussed in Sec.~\ref{sec:glide-DIII}. 
In this section, we clarify the topological invariant protecting the surface states. 
In addition to the glide symmetry, we take the mirror symmetry into account. 
The Weyl charge, mirror Chern number, glide winding number, and rotation winding number 
are discussed below.

\subsection{Chiral Majorana arc in Weyl B-phase}\label{sec:Weyl}

The TRS broken B-phase identified as a Weyl superconducting state~\cite{Yanase_UPt3_Weyl,Goswami}  
hosts surface Majorana arcs, analogous to Fermi arcs in Weyl 
semimetals~\cite{Murakami,Wan-Vishwanath,Burkov-Balents,Xu,Lv,Yang,Huang}. 
The existence of Majorana arcs is ensured by the topological Weyl charge  
%
%$
\begin{equation}
q_i = \frac{1}{2\pi} \oint_{S} {\rm d}{\bm k} \vec{F}({\bm k}), 
\end{equation}
%$
which is nothing but the monopole of Berry flux, 
\begin{align}
F_{i}({\bm k}) = -i \varepsilon^{ijk}\sum_{E_n(\k)<0} \partial_{k_j} \langle u_n(\k)| \partial_{k_k} u_n(\k)\rangle. 
%\nonumber \\
  \label{Berry_flux}
\end{align}
A wave function and energy of Bogoliubov quasiparticles are denoted by $|u_n(\k)\rangle$ and $E_n(\k)$, respectively. 
A nontrivial Weyl charge protects the Weyl point node in the bulk excitation spectrum. 
Indeed, the B-phase of UPt$_3$ is a point nodal SC compatible with Blount's theorem~\cite{Blount,Kobayashi-Sato} 
when the $p$-wave and $d$-wave order parameters are appropriately taken into account~\cite{Norman1995,Yanase_UPt3_Weyl}. 
Although the purely $f$-wave state has a nodal line at $k_z=0$~\cite{Sauls}, it is an accidental node 
removed by symmetry-preserving perturbation.  
In accordance with the bulk-boundary correspondence, the Majorana arcs appear on the surface and terminate 
at the projection of Weyl point nodes illustrated by green circles in 
Figs.~\ref{eGFSedgestate} and \ref{ZFSedgestate}.

Interestingly, the position of Weyl nodes is tunable.  In the $E_{\rm 2u}$ scenario for UPt$_3$~\cite{Sauls}, 
the parameter $\eta$ smoothly changes from $\infty$ to $0$ in the B-phase 
by decreasing the temperature and/or increasing the magnetic field (see Fig.~\ref{phasediagram} and Table~I). 
Then, the pair creation, pair annihilation, and coalescence of Weyl nodes occur as a consequence of 
the $p$-$f$ mixing in the order parameter~\cite{Yanase_UPt3_Weyl}. 
Accordingly, the projection of Weyl nodes moves as illustrated in Figs.~\ref{eGFSedgestate}(b)-(d) 
and \ref{ZFSedgestate}(b)-(d). The Majorana arcs follow the Weyl nodes.

In the generic $E_{\rm 2u}$-state studied in this paper, the Weyl nodes are purely protected by topology, 
and any crystal symmetry is not needed. 
Therefore, the positions of Weyl nodes are not constrained by any symmetry. 
Although the Weyl nodes are pinned at the poles of FS in the purely $f$-wave $E_{\rm 2u}$-state~\cite{Goswami}, 
that is an accidental result. 
In another candidate of Weyl SC URu$_2$Si$_2$, the 3D $d_{xz} \pm id_{yz}$-wave 
superconductivity has been revealed by experiments.~\cite{Kasahara2007,Yano2008,Kittaka2016,Yamashita2015} 
Then, the Weyl nodes are pinned and the traveling of Weyl nodes does not occur, 
in contrast to UPt$_3$.

\begin{figure}[htbp]
\begin{center}
\includegraphics[width=90mm]{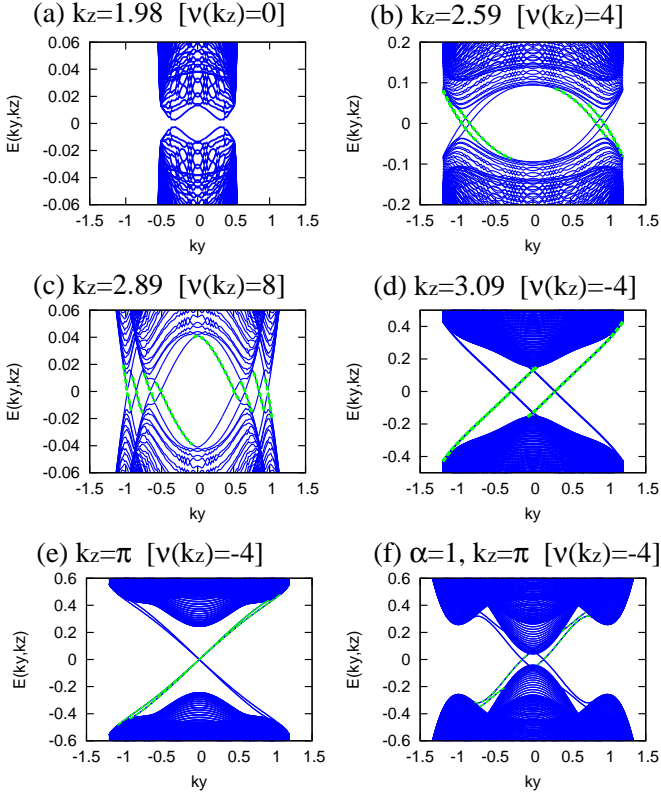}
\caption{(Color online) Energy spectrum on the (100)-surface in the B-phase ($\eta=0.6$). 
Surface and bulk quasiparticle states on slices of BZ at $k_z=$ constant planes are shown. 
The surface states are emphasized by green lines. 
The $k_z$-dependent Chern number is shown in each panel. 
(a)-(e) Parameters are the same as Fig.~\ref{ZFSedgestate}(b). 
(f) $\alpha=1$ while the others are the same as (e). 
The surface states are almost two-fold degenerate in (d) and (f) and four-fold degenerate in (e). 
Comparison of (e) and (f) reveals that the four-fold degeneracy in the absence of the SOC is lifted by the SOC. 
} 
\label{ZFSedgestate0.6}
\end{center}
\end{figure}

Here the number of Majorana arcs is verified by calculating the Chern number~\cite{Thouless,Kohmoto,Fukui} 
of effective 2D models on $k_z=$ constant planes, 
\begin{align} 
&
\nu(k_z) = \frac{1}{2\pi} \int {\rm d}{\bm k}_\parallel F_{z}({\bm k}), 
\label{Chern}
\end{align}
that is, a $k_z$-dependent Chern number of class A. 
The Chern number indicates the number of chiral surface modes. 
In Weyl SCs, the Chern number may change at a gapless $k_z=$ constant plane hosting Weyl nodes. 
Therefore, the zero energy surface states form arcs terminating at the projection of Weyl nodes. 
For parameters reproducing the $A$-FSs, the Chern number changes  
$\nu(k_z)= 0 \rightarrow 4 \rightarrow 8 \rightarrow -4$ with increasing $k_z$ from $0$ to $\pi$, 
while $\nu(k_z)= 0 \rightarrow 4 \rightarrow 0$ for the $\Gamma$-FS~\cite{Yanase_UPt3_Weyl}. 
The bulk-boundary correspondence is confirmed by showing the surface spectrum on the $k_z=$ constant lines 
in Fig.~\ref{ZFSedgestate0.6}. The number of chiral modes coincides with the $k_z$-dependent Chern number. 
We also observe the sign reversal of chirality in accordance with the sign change of the Chern number.

Finally, we discuss the Weyl superconducting phase in the phase diagram illustrated in Fig.~\ref{phasediagram}.
Because the TRS has to be broken in Weyl SCs, the A- and C-phases are non-Weyl superconducting states. 
Furthermore, the B-phase in the vicinity of the A-B and B-C phase boundaries is also 
non-Weyl state because the gap closing is required for the topological transition. 
Therefore, the transition from the non-Weyl state to the Weyl state occurs in the B-phase. 
The shaded region in Fig.~\ref{phasediagram} schematically illustrates the Weyl superconducting phase.

\subsection{Majorana cone and mirror Chern number}\label{sec:mirror}

\begin{figure}[htbp]
\begin{center}
\includegraphics[width=85mm]{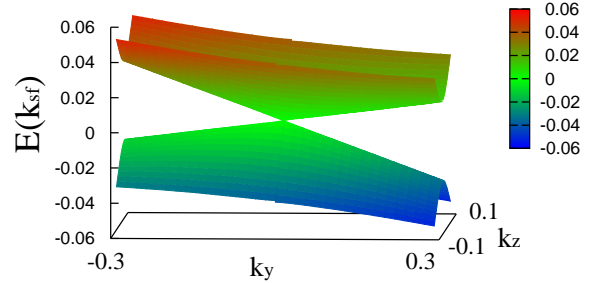}
\caption{(Color online) Tilted Majorana cone at $\k_{\rm sf}=(0,0)$ in the B-phase ($\eta=0.5$). 
Parameters are $(t,t_z,t',\alpha,\mu,\Delta,\delta)=(1,4,1,0,16,4,0.02)$ reproducing a $\Gamma$-FS. 
} 
\label{mirror-cone}
\end{center}
\end{figure}

Next we discuss the surface state around ${\bm k}_{\rm sf} = (0,0)$, which is observed 
in all the A-, B-, and C-phases (Fig.~\ref{eGFSedgestate}). 
In the TRS broken B-phase, the spectrum resembles a tilted Majorana cone as shown in Fig.~\ref{mirror-cone}, 
although the cone is not tilted in the A- and C-phases. 
Naturally, the $\Gamma$-FS and $K$-FS are considered in this subsection. 

We may understand the topological protection by implementing the crystal mirror reflection symmetry 
with respect to the $xy$-plane. 
Mirror reflection operator for the normal part Hamiltonian is, 
\begin{align}
& M^{xy}(k_z) = i s_z V_{\sigma}(k_z). 
  \label{Mirror-operator}
\end{align}
The mirror reflection symmetry is equivalent to the product of inversion symmetry and  screw symmetry, 
that is, $M^{xy}(k_z)=I S_\pi^{z}(k_z)$. 
The nonsymmorphic screw symmetry $S_\pi^{z} = \{R_\pi^{z}|\frac{z}{2}\}$ involves half translation along 
the $z$-axis, and therefore, the screw operator $S_\pi^{z}(k_z)$ is $k_z$-dependent. 
Thus, the mirror operator is also $k_z$-dependent, and we have $M^{xy}(\pi) = i s_z \sigma_z$ while $M^{xy}(0) = i s_z $. 
This momentum dependence of $M^{xy}(k_z)$ may yield the unusual line node in nonsymmorphic odd-parity 
SCs~\cite{Kobayashi-Yanase-Sato}, a counterexample of Blount's theorem.

The normal part Hamiltonian is invariant under the mirror reflection symmetry
\begin{align}
&
M^{xy}(k_z) \tilde{H}_0(\k) M^{xy}(k_z)^{-1} = \tilde{H}_0(k_x,k_y,-k_z), 
\end{align}
and the order parameter is mirror-odd irrespective of $\eta$, 
\begin{align}
&& 
M^{xy}(k_z) \tilde{\Delta}(\k) M^{xy}(-k_z)^{\rm T} = -\tilde{\Delta}(k_x,k_y,-k_z). 
\end{align}
Thus, the BdG Hamiltonian respects mirror reflection symmetry, 
\begin{align}
&& \hspace{-0mm}
M^{xy}_{\rm BdG}(k_z) \tilde{H}_{\rm BdG}(\k) M^{xy}_{\rm BdG}(k_z)^{-1} = \tilde{H}_{\rm BdG}(k_x,k_y,-k_z), 
\end{align}
by defining the operator in the Nambu space, 
\begin{align}
M^{xy}_{\rm BdG}(k_z) 
&=
\left(
\begin{array}{cc}
M^{xy}(k_z)  & 0 \\
0 & -M^{xy}(-k_z)^* \\
\end{array}
\right)_{\tau}
\\ 
&= M^{xy}(k_z) \otimes \tau_0.
\end{align}

According to the $K$-theory for topological crystalline insulators and SCs~\cite{Shiozaki2014}, 
the effective 2D Hamiltonian at mirror invariant planes, namely, $k_z=0$ and $\pi$, 
is specified by a topological index of class D, $\mathbb{Z} \oplus \mathbb{Z}$, in the TRS broken B-phase. 
This is ensured by the algebra
$\left[M^{xy}_{\rm BdG}(0)\right]^2 = \left[M^{xy}_{\rm BdG}(\pi)\right]^2=-1$  
and $\{M^{xy}_{\rm BdG}(0),C \} = \{M^{xy}_{\rm BdG}(\pi),C\} = 0$.  
One of the two integer topological invariants is nothing but the Chern number $\nu(k_z)$ introduced 
in Sec.~\ref{sec:Weyl}. 
The other is the mirror Chern number, $\nu_{\rm M}^{\Gamma_z} \in \mathbb{Z}$ $\,\,$($\Gamma_z \equiv 0$ or $\pi$), 
which is defined below by using the mirror reflection symmetry~\cite{Ueno-Sato,Yoshida2015}. 
In the TRS invariant A- and C-phases, the Chern number must be zero, and the mirror Chern number 
is naturally the $\mathbb{Z}$ topological index of class DIII appearing in Ref.~\onlinecite{Shiozaki2014}.

The commutation relation, 
$\left[M^{xy}_{\rm BdG}(\Gamma_z), \tilde{H}_{\rm BdG}(\k_\parallel,\Gamma_z)\right] =0$, ensures that 
the BdG Hamiltonian is block-diagonalized at mirror invariant planes on the basis diagonalizing 
$M^{xy}_{\rm BdG}(\Gamma_z)$. In other words, the BdG Hamiltonian is decomposed into two mirror-subsectors 
with mirror eigenvalues $\pm i$, 
\begin{align}
\tilde{H}_{\rm BdG}(\k_\parallel,\Gamma_z) = \tilde{H}^{\Gamma_z}_{i}(\k_\parallel) \oplus \tilde{H}^{\Gamma_z}_{-i}(\k_\parallel). 
\end{align} 
The PHS is preserved in the mirror-subsector, because of 
$[M^{xy}_{\rm BdG}(\Gamma_z)]^2 = -1$ and $\{M^{xy}_{\rm BdG}(\Gamma_z), C\}=0$. On the other hand, the TRS 
is not preserved even in the TRS invariant A- and C-phases since $[M^{xy}_{\rm BdG}(\Gamma_z), T]=0$. 
Thus, the symmetry of the mirror-subsector is class D irrespective of $\eta$, and 
the Chern number of mirror-subsector Hamiltonian given by 
\begin{align} 
&
\nu^{\Gamma_z}_{\pm i} = \frac{1}{2\pi} \int {\rm d}{\bm k}_\parallel F_{z, \pm i}^{\Gamma_z}({\bm k}_\parallel), 
\label{mirror-Chern}
\end{align}
may be nontrivial. 
Here, $F_{z, \pm i}^{\Gamma_z}({\bm k}_\parallel)$ is the Berry curvature of $\tilde{H}^{\Gamma_z}_{\pm i}(\k_\parallel)$. 
The mirror Chern number is defined by 
\begin{align}
\nu_{\rm M}^{\Gamma_z} = \nu^{\Gamma_z}_{+i} \, - \, \nu^{\Gamma_z}_{-i}, 
\end{align} 
while the total Chern number is given by $\nu(\Gamma_z) = \nu^{\Gamma_z}_{+i} \, + \, \nu^{\Gamma_z}_{-i}$. 

\subsubsection{Mirror Chern number at $k_z=0$}

Later we show that the mirror Chern number at $k_z= \pi$ has to vanish owing to the constraint 
by glide symmetry. On the other hand, the mirror Chern number may be nontrivial at $k_z =0$, 
and the surface states around $\k_{\rm sf}=(0,0)$ are indeed protected by the mirror Chern number.

Because we have $M^{xy}_{\rm BdG}(0) = i s_z \sigma_0 \tau_0$, the mirror-subsector Hamiltonian 
$\tilde{H}^0_{\pm i}(\k_\parallel)$ is equivalent to the spin sector for $s=\uparrow$ and $\downarrow$, 
respectively. Thus, we obtain 
\begin{align}
\tilde{H}^0_{\pm i}(\k_\parallel)
&=
\left(
\begin{array}{cc}
\hat{h}_{\pm i}(\k_\parallel)  & \hat{\Delta}_{\pm i}(\k_\parallel) \\
\hat{\Delta}_{\pm i}(\k_\parallel)^\dag & - \hat{h}_{\pm i}(-\k_\parallel)^{T} \\
\end{array}
\right), 
\end{align}
with 
\begin{align}
\hspace{-3mm}
\hat{h}_{\pm i}(\k_\parallel) 
&=
%\nonumber \\ && \hspace{3mm}
\left(
\begin{array}{cc}
\varepsilon(\k_\parallel) \pm \alpha g(\k_\parallel)  & \tilde{a}(\k_\parallel)  \\
\tilde{a}(\k_\parallel)^*  & \varepsilon(\k_\parallel) \mp \alpha g(\k_\parallel)  \\
\end{array}
\right), 
\end{align}
and 
\begin{align}
\hat{\Delta}_{\pm i}(\k_\parallel) 
&= 
-(\eta \pm 1) \Delta_{\rm p} \left[p_{x}(\k_\parallel) \pm i p_y(\k_\parallel)  \right] \sigma_0.
\label{subsector-OP_0}
\end{align}
We denoted $A(\k_\parallel)=A(\k_\parallel,0)$ and $\Delta_{\rm p} = \delta \Delta/\sqrt{1+\eta^2}$. 
It turns out that the mirror-subsector Hamiltonian is equivalent to the BdG Hamiltonian of 
a two-band chiral $p$-wave SC. 
In our model for the $\Gamma$-FS, only one band crosses the Fermi level, 
and we obtain $\nu^0_{\pm i}= \pm 1$. 
The sign of Chern number is opposite between the two mirror-subsectors because of 
the opposite chirality of $p$-wave order parameter [see Eq.~(\ref{subsector-OP_0})]. 
Therefore, the total Chern number of the 2D BdG Hamiltonian is zero, 
$\nu(0) = \nu^0_{+i} + \nu^0_{-i} = 0$, even in the TRS broken B-phase. 
On the other hand, the mirror Chern number is nontrivial, 
\begin{align}
\nu_{\rm M}^0= 2. 
\label{mirror_Chern_Gamma}
\end{align}

We now understand that the (tilted-) Majorana cone in Fig.~\ref{mirror-cone} is the topological surface states 
ensured by the bulk-boundary correspondence. 
Since the chirality of Majorana modes corresponding to $\nu^0_{\pm i}= \pm 1$ is opposite between 
two mirror-subsectors, the (tilted) helical mode appears at $k_z=0$, and the helical mode is gapped 
at $k_z \ne 0$, implying the Majorana cone.

Finally, we comment on the multiband effect. 
Although Eq.~(\ref{mirror_Chern_Gamma}) is obtained for a hole $\Gamma$-FS, we obtain $\nu_{\rm M}^0= -2$ for 
a electron $\Gamma$-FS consistent with UPt$_3$~\cite{Taillefer1988,Kimura_UPt3,McMullan,Nomoto}. 
Because the mirror Chern number is additive, we will obtain the mirror Chern number $\nu_{\rm M}^0= -6$ from 
three $\Gamma$-FSs. Then, the surface states form a (tilted-) Majorana cone at ${\bm k}_{\rm sf} = (0,0)$ 
and two cones away from the $\Gamma$-point, ${\bm k}_{\rm sf} = (\pm k_y^0,0)$. 
Although the $K$-FSs have also been predicted by band structure calculations~\cite{Taillefer1988,Nomoto}, 
the existence of them is still under debate~\cite{McMullan}. 
The $K$-FSs also give nontrivial mirror Chern number $\nu_{\rm M}^0= -8$ if they exist. 
Then, the mirror Chern number is $\nu_{\rm M}^0= -14$ by taking into account all the FSs. 
In any case, $\nu_{\rm M}^0 \in  4 \Z +2$ indicates the existence of Majorana cone at ${\bm k}_{\rm sf} = (0,0)$.

\subsubsection{Vanishing mirror Chern number at $k_z=\pi$}

We here show that the mirror Chern number at $k_z=\pi$ must be trivial owing to the glide symmetry, 
namely, 
\begin{align}
\nu_{\rm M}^\pi= 0. 
\label{mirror-Chern-pi}
\end{align}
First, we consider the glide invariant A- and C-phases. 
The glide symmetry is also preserved in the mirror-subsectors at $k_z =\pi$ because of 
$\left[M^{xy}_{\rm BdG}(\pi), G^{xz}_{\rm BdG}(\pi)\right] = 0$, although 
$\left\{M^{xy}_{\rm BdG}(0), G^{xz}_{\rm BdG}(0)\right\} = 0$ indicates the broken glide-symmetry in 
the mirror-subsectors at $k_z=0$.  
Then, we can prove the relation for Berry curvature, 
\begin{align}
F_{z, \pm i}^{\pi}(k_x,k_y) = - F_{z, \pm i}^{\pi}(k_x,-k_y). 
\end{align}
Integration over the $(k_x,k_y)$ plane ends up vanishing Chern number, $\nu^{\pi}_{\pm i} =0$, and thus, 
the mirror Chern number also vanishes. 
In the B-phase, the glide symmetry is spontaneously broken. However, considering the magnetic-glide 
symmetry $T G^{xz}_{\rm BdG}(\pi)$, we can show the relation
\begin{align}
F_{z, \pm i}^{\pi}(k_x,k_y) = F_{z, \mp i}^{\pi}(-k_x,k_y), 
\end{align}
which leads to $\nu^{\pi}_{i} = \nu^{\pi}_{-i} $. 
Therefore, the mirror Chern number at $k_z = \pi$ vanishes in the B-phase as well.

The trivial mirror Chern number is confirmed in our model as follows. 
Using the mirror reflection operator $M^{xy}_{\rm BdG}(\pi) = i s_z \sigma_z \tau_0$, 
we obtain the mirror-subsector Hamiltonian respecting the PHS, 
\begin{align}
\tilde{H}^\pi_{\pm i}(\k_\parallel)
&=
\left(
\begin{array}{cc}
\hat{h}_{\pm i}(\k_\parallel)  & \hat{\Delta}_{\pm i}(\k_\parallel) \\
\hat{\Delta}_{\pm i}(\k_\parallel)^\dag & - \hat{h}_{\pm i}(-\k_\parallel)^{T} \\
\end{array}
\right). 
\end{align}
The normal part is given by 
\begin{align}
\hspace{-3mm}
\hat{h}_{\pm i}(\k_\parallel) 
&=
%\nonumber \\ && \hspace{3mm}
%\left(
%\begin{array}{cc}
\left[\varepsilon(\k_\parallel) \pm \alpha g(\k_\parallel)\right] \sigma_0. 
%& 0 \\
%0 & \varepsilon(\k_\parallel) \pm \alpha g(\k_\parallel)  \\
%\end{array}
%\right), 
\end{align}
For instance, the order parameter part is 
\begin{align}
\hat{\Delta}_{\pm i}(\k_\parallel) 
&= 
\Delta \times 
\nonumber \\
& \hspace{-12mm}
\left(
\begin{array}{cc}
\mp \delta \left[p_{x}(\k_\parallel) \pm i p_y(\k_\parallel)  \right] 
& 
\tilde{f}_{(x^2-y^2)z}(\k_\parallel) + i \tilde{d}_{yz}(\k_\parallel) 
\\
\tilde{f}_{(x^2-y^2)z}(\k_\parallel)^* - i \tilde{d}_{yz}(\k_\parallel)^* 
& 
\pm \delta \left[p_{x}(\k_\parallel) \mp i p_y(\k_\parallel)  \right] 
\end{array}
\right), 
\nonumber \\
\label{subsector-OP-pi}
\end{align}
in the C-phase. 
When the $d+f$-wave component is dominant as we assume in this paper, the $p$-wave component can be 
adiabatically reduced to zero without closing the gap. Then, it turns out that the Chern number of 
mirror-subsectors is trivial because the phase winding of 
$\tilde{f}_{(x^2-y^2)z}(\k_\parallel) \pm i \tilde{d}_{yz}(\k_\parallel)$ along the FS is zero. 
Even when the $p$-wave component is dominant, the Chern number vanishes 
because the chirality of gap function $p_{x}(\k_\parallel) \pm i p_y(\k_\parallel)$ is opposite 
between the pseudospin up and down Cooper pairs. 
Thus, we obtain $\nu^\pi_{\pm i} =0$ and $\nu_{\rm M}^\pi =0$ in the C-phase. 
It is straightforward to show $\nu^\pi_{\pm i} =0$ in the A-phase as well.
In the B-phase we have obtained the nontrivial Chern number $\nu(\pi)=-4$ for the $A$-FSs. However, 
the mirror Chern number remains trivial, because $\nu^\pi_{\pm i}=-2$. 
We have numerically confirmed Eq.~(\ref{mirror-Chern-pi}) in the entire A-, B- and C-phases for all the FSs.

\subsection{Majorana flat band and glide winding number}\label{sec:glide-AIII}

As shown in Figs.~\ref{eGFSedgestate}(d) and (e) and Figs.~\ref{ZFSedgestate}(d) and (e), 
the zero energy surface flat band appears in the A-phase and in a ``half'' of the B-phase ($|\eta| > 1$). 
We here show that the flat band is topologically protected by the glide winding number. 
Below we first demonstrate the topological protection in the A-phase, and later investigate the B-phase.

\subsubsection{A-phase}

Let's consider the glide invariant plane $k_y=0$ in the A-phase. 
The glide winding number is defined for 1D models $\tilde{H}_{\rm BdG}(k_x,0,k_z)$ parametrized by $k_z$. 
The 1D models do not respect TRS and PHS unless $(0,0,k_z)$ is a time-reversal invariant momentum. 
On the other hand, the combined chiral symmetry, $\Gamma = i T_{\rm BdG} \, C$, is preserved. 
Thus, the winding number of 1D AIII class can be defined. However, it is obtained to be zero.

The nontrivial winding number is obtained by implementing the 
glide symmetry, which has been represented by Eq.~(\ref{glide-symmetry}). 
The glide symmetry ensures the sector decomposition, 
\begin{align}
\tilde{H}_{\rm BdG}(k_x,0,k_z) = \tilde{H}_{\lambda_+}(k_x,k_z) \oplus \tilde{H}_{\lambda_-}(k_x,k_z),
\end{align} 
for eigenvalues $\lambda_{\pm} = \pm i e^{- i k_z/2}$ of the glide operator. 
The chiral symmetry is preserved in the glide-subsector Hamiltonian 
$\tilde{H}_{\lambda_\pm}(k_x,k_z)$ because $\left[ \Gamma, G^{xz}_{\rm BdG}(k_z)\right]=0$ 
in the A-phase~\cite{Comment1}. 
Now we have two winding numbers of AIII class, ${\cal \omega}_{\rm G}(+, k_z)$ and ${\cal \omega}_{\rm G}(-, k_z)$, 
which correspond to the $\mathbb{Z} \oplus \mathbb{Z}$ topological index of 1D AIII class 
with $U_+$ crystal symmetry~\cite{Shiozaki2014}.

We here estimate the winding numbers by analyzing the original BdG Hamiltonian $\hat{H}_{\rm BdG}({\bm k})$, 
instead of $\tilde{H}_{\rm BdG}({\bm k})$.
The periodicity along the $k_x$-axis is satisfied in $\hat{H}_{\rm BdG}({\bm k})$, and  
the unitary transformation (\ref{unitary_transformation}) does not alter the winding number, 
since $\left[\Gamma,U(\k) \right]=0$. 
The glide operator for the original BdG Hamiltonian $\hat{H}_{\rm BdG}({\bm k})$ is 
$G^{xz}_{\rm BdG} = i s_y \sigma_x \tau_0 e^{- i k_z/2}$ in the A-phase while $G^{xz}_{\rm BdG} = i s_y \sigma_x \tau_z e^{- i k_z/2}$ 
in the C-phase.

In the A-phase, the glide-subsectors of $\hat{H}_{\rm BdG}({\bm k})$ are, 
\begin{align}
\hat{H}_{\lambda_\pm}(k_x,k_z) =& \, \varepsilon^{k_z}(k_x) \sigma_0 \tau_z \pm a^{k_z}(k_x) \sigma_z \tau_z 
%\nonumber \\ &
+ \alpha g(k_x) \sigma_y \tau_0
\nonumber \\ &
-\Delta \delta p_x(k_x) \sigma_0 \tau_x 
%\nonumber \\ &
\pm \Delta d_{xz}^{\, k_z}(k_x) \sigma_y \tau_y. 
\label{glide_subsector_AIII}
\end{align} 
The chiral symmetry is confirmed by $\left\{\Gamma_{\rm s}, \hat{H}_{\lambda_\pm}(k_x, k_z)\right\}=0$, 
where $\Gamma_{\rm s} = \sigma_z \tau_y$ is the chiral operator in the subsector space. 
Thus, we obtain the off-diagonal form 
\begin{align}
&\hspace{-0mm}
U_{\Gamma_{\rm s}} \hat{H}_{\lambda_\pm}(k_x, k_z) U_{\Gamma_{\rm s}}^\dagger =
\left(
\begin{array}{cc}
0 & \hat{q}_{\pm}(k_x,k_z) \\
\hat{q}_{\pm}^\dagger(k_x,k_z) & 0 \\
\end{array}
\right), 
\label{glide-off-diagonal}
\end{align}
by choosing the basis diagonalizing the chiral operator.  
From Eq.~(\ref{glide_subsector_AIII}), we obtain 
\begin{align}
q_{\pm}(k_x, k_z)  = &
i\varepsilon^{k_z}(k_x) \sigma_z \pm i a^{k_z}(k_x) \sigma_0 +  \Delta \delta p_x(k_x) \sigma_0 
\nonumber \\ &
+ [\alpha g(k_x) \pm \Delta d_{xz}^{\, k_z}(k_x)] \sigma_y, 
%\nonumber \\ &
%\left(
%\begin{array}{cc}
%i[\varepsilon^{k_z}(k_x) \pm a^{k_z}(k_x)] + \Delta \delta p_x(k_x) & -i [\alpha g(k_x) \pm \Delta d_{xz}^{\, k_z}(k_x)] \\
%i [\alpha g(k_x) \mp \Delta d_{xz}^{\, k_z}(k_x)]  & -i[\varepsilon^{k_z}(k_x) \mp a^{k_z}(k_x)] + \Delta \delta p_x(k_x) \\
%\end{array}
%\right), 
%\nonumber \\ 
\end{align}
for the $\lambda_\pm$ glide-subsector, respectively. 
We used abbreviations, $A^{k_z}(k_x) =A(k_x,0,k_z)$. 

Now the winding number of glide-subsectors given by 
\begin{align}
\label{glide-winding}
\hspace{-0mm} 
{\cal \omega}_{\rm G}(\pm, k_z) = \frac{1}{4\pi i} \int_{0}^{4\pi} dk_x {\rm Tr}
& \Big[ \hat{q}_{\pm}(k_x,k_z)^{-1}\partial_{k_x}\hat{q}_{\pm}(k_x,k_z) 
\nonumber \\ & \hspace{-5mm}
- \hat{q}_{\pm}^\dagger(k_x,k_z)^{-1}\partial_{k_x}\hat{q}_{\pm}^\dagger(k_x,k_z) \Big], 
\end{align}
is calculated. 
By adiabatically reducing $\alpha g(k_x) \rightarrow 0$ and $d_{xz}^{\, k_z}(k_x) \rightarrow 0$ 
without closing the excitation gap, we obtain the winding number as 
\begin{align}
& {\cal \omega}_{\rm G}(\pm, k_z) 
\nonumber \\ & =
\begin{cases}
\mp 1 & [\varepsilon({\bm 0},k_z) + a({\bm 0},k_z) > 0 > \varepsilon({\bm 0},k_z) - a({\bm 0},k_z)] \\
0 & [{\rm otherwise}]
\end{cases}, 
\nonumber \\
\label{glide-windingA}
\end{align}
for $t>0$, $t'>0$ and $\Delta \delta >0$. 
This means that the $\lambda_\pm$ glide-subsectors of $\hat{H}_{\rm BdG}({\bm k})$ [and equivalently 
the subsectors of the periodic BdG Hamiltonian $\tilde{H}_{\rm BdG}({\bm k})$]
are topologically characterized by the glide-winding number 
${\cal \omega}_{\rm G}(\pm, k_z) = \mp 1$, when the condition 
$\varepsilon({\bm 0},k_z) + a({\bm 0},k_z) > 0 > \varepsilon({\bm 0},k_z) - a({\bm 0},k_z)$ is satisfied. 
This condition is equivalent to the number of FSs (per Kramers pairs) is odd.

In Figs.~\ref{eGFSedgestate}(e) and \ref{ZFSedgestate}(e), the flat band appears 
on the $k_y=0$ line of surface BZ where only one FS is projected. 
The nontrivial glide-winding number demonstrated above protects this Majorana flat band. 
The zero energy states are two-fold degenerate in accordance with the bulk-boundary correspondence. 
One comes from the $\lambda_{+} = i e^{- i k_z/2}$ glide-subsector and the other comes from 
the $\lambda_{-} = - i e^{- i k_z/2}$ glide-subsector. 
Note that the flat band is robust against the multiband effect. We find that the glide-winding number 
of $K$-FSs is zero. Taking into account three $\Gamma$-FSs, we will have glide-winding number 
${\cal \omega}_{\rm G}(\pm, 0) = \mp 3$, or $\mp 1$, or $\pm 1$, or $\pm 3$ 
depending on the sign of order parameter. In any case, the glide-winding number is nontrivial.

\subsubsection{B-phase}

The glide-subsector is no longer well-defined in the B-phase, because the glide symmetry 
is spontaneously broken. However, the glide-winding number is well-defined 
by the magnetic-glide symmetry $G^{xz}_{\rm BdG} T$ preserved in the B-phase. 
Then, the glide-winding number is given by 
\begin{align}
{\cal \omega}_{\rm G}(k_z) = \frac{i}{4\pi} \int_{0}^{4\pi} dk_x {\rm Tr} & \Big[
\Gamma_{\rm G} \tilde{H}_{\rm BdG}(k_x,0,k_z)^{-1} 
\nonumber \\
& \times \partial_{k_x} \tilde{H}_{\rm BdG}(k_x,0,k_z) \Big],
\label{glide-winding2}
\end{align}
where $\Gamma_{\rm G} = e^{i \phi} G^{xz}_{\rm BdG}(k_z) T_{\rm BdG} C $ is the glide-chiral operator
with  $\Gamma_{\rm G}^2=1$. 

\begin{figure}[htbp]
\begin{center}
\includegraphics[width=80mm]{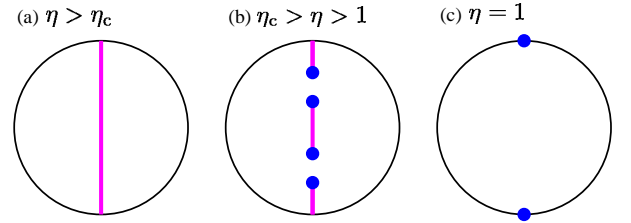}
\caption{(Color online) 
Illustration of the Majorana flat band  
(a) in the A-phase and non-Weyl B-phase ($\eta > \eta_{\rm c}$), 
(b) in the Weyl B-phase  ($\eta_{\rm c} > \eta > 1$), and  
(c) at the critical point ($\eta = 1$). 
Thick solid (purple) lines show the Majorana flat band. Thin lines illustrate the projection of a $\Gamma$-FS onto 
the (100)-surface BZ. The closed (blue) circles indicate projections of Weyl point nodes. 
(a), (b), and (c) correspond to the numerical results in Figs.~\ref{eGFSedgestate}(e), (d), and (c), respectively. 
} 
\label{schematic-glide}
\end{center}
\end{figure}

In the A-phase, Eq.~(\ref{glide-winding2}) is reduced to 
\begin{align}
{\cal \omega}_{\rm G}(k_z) ={\cal \omega}_{\rm G}(+,k_z) - {\cal \omega}_{\rm G}(-,k_z).
\end{align}
Thus, we obtain ${\cal \omega}_{\rm G}(k_z)=-2$ in the A-phase. 
The nontrivial glide-winding number is robust as long as the gap is finite. 
Therefore, the Majorana flat band appears in the B-phase under the condition (\ref{glide-windingA}), 
when the parameter $|\eta|$ is large [see Fig.~\ref{schematic-glide}(a)]. 
When $|\eta|$ is decreased from infinity, the pair creation of Weyl nodes occurs in the bulk BZ 
on the $k_y=0$ plane~\cite{Yanase_UPt3_Weyl}. Then, a part of the Majorana flat band disappears 
in between the pair of projected Weyl points [see Fig.~\ref{schematic-glide}(b)]. 
Therefore, the projected Weyl points are end points not only of the Majorana arc but also of the Majorana flat band. 
This feature has been shown in Figs.~\ref{eGFSedgestate}(d) and \ref{ZFSedgestate}(d).

At $|\eta| =1$, a pair of Weyl nodes is annihilated on the $\k = (k_x,0,0)$ line, and other Weyl nodes 
coalesce on the poles of FSs~\cite{Yanase_UPt3_Weyl}. 
Then, the Majorana flat band completely disappears [Fig.~\ref{schematic-glide}(c)]. 
The fate of the Majorana flat band in the B-phase is schematically illustrated in Fig.~\ref{schematic-glide},  
and shown in Figs.~\ref{eGFSedgestate} and \ref{ZFSedgestate} by the numerical diagonalization 
of the BdG Hamiltonian.

\subsection{Symmetry constraint on winding numbers}\label{sec:symmetry}

The crystal symmetries preserved on the (100)-surface are as follows. 
\begin{itemize}
\item Mirror symmetry $M^{xy}$.
\item Glide symmetry $G^{xz}$. 
\item $\pi$-rotation symmetry $R^x$. 
\end{itemize}
The $\pi$-rotation is given by the product of mirror and glide operations.  

In addition to the glide-winding number studied in Sec.~\ref{sec:glide-AIII}, 
we can define the mirror-winding number~\cite{Tsutsumi_UPt3} and the rotation-winding number~\cite{Mizushima_3He} 
in the same manner. 
They are given by 
\begin{align}
{\cal \omega}_{\rm M}^{\Gamma_z}(k_y) = \frac{i}{4\pi} \int_{0}^{4\pi} dk_x {\rm Tr} & \Big[
\Gamma_{\rm M}(\Gamma_z) \tilde{H}_{\rm BdG}(k_x,k_y,\Gamma_z)^{-1} 
\nonumber \\
& \times \partial_{k_x} \tilde{H}_{\rm BdG}(k_x,k_y,\Gamma_z) \Big],
\label{mirror-winding}
\end{align}
and 
\begin{align}
{\cal \omega}_{\rm R}^{\Gamma_z} = \frac{i}{4\pi} \int_{0}^{4\pi} dk_x {\rm Tr} & \Big[
\Gamma_{\rm R} \tilde{H}_{\rm BdG}(k_x,0,\Gamma_z)^{-1} 
\nonumber \\
& \times \partial_{k_x} \tilde{H}_{\rm BdG}(k_x,0,\Gamma_z) \Big]. 
\label{rotation-winding}
\end{align}
$\Gamma_{\rm M}(\Gamma_z) = e^{i\theta} M^{xy}_{\rm BdG}(\Gamma_z) \Gamma$ and 
$\Gamma_{\rm R} = e^{i\theta'}R^{x}_{\rm BdG} \Gamma$ are mirror-chiral operator and rotation-chiral operator, respectively. 
The phase factors $e^{i\theta}$ and $e^{i\theta'}$ are chosen so that $\Gamma_{\rm M}(\Gamma_z)^2 =\Gamma_{\rm R}^2=1$. 
The mirror-winding number is defined on the mirror invariant planes at $k_z = \Gamma_z = 0, \pi$ 
and $k_y$-dependent. 
On the other hand, the rotation-winding number is defined on the rotation invariant lines. 
The mirror-winding number is defined only in the TRS invariant A- and C-phases, since the mirror-chiral symmetry is broken 
in the TRS broken B-phase. 

From the algebra of symmetry operations we can prove that most of the winding numbers vanish. 
The proof relies on the fact that the winding number disappears when any unitary symmetry preserved 
on the surface anti-commutes with the chiral operator, $\left\{ U, \Gamma_{V} \right\} =0$. 
This fact, ${\cal \omega}_{\rm V}=0$, is understood by 
\begin{align}
{\cal \omega}_{\rm V} & = \frac{i}{4\pi} \int_{0}^{4\pi} dk_x {\rm Tr} \Big[U 
\Gamma_{V} \tilde{H}_{\rm 1D}(k_x)^{-1} \partial_{k_x} \tilde{H}_{\rm 1D}(k_x) U^\dag \Big] 
\nonumber \\ & = \frac{i}{4\pi} \int_{0}^{4\pi} dk_x {\rm Tr} \Big[
\left(-\Gamma_{V}\right) \tilde{H}_{\rm 1D}(k_x)^{-1} \partial_{k_x} \tilde{H}_{\rm 1D}(k_x) \Big] 
\nonumber \\ & = - {\cal \omega}_{\rm V}.  
\end{align}
Furthermore, the TRS has to satisfy $[T, \Gamma_V]=0$ when the winding number is nontrivial. 
All of the mirror, glide, and rotation symmetries are preserved at the rotation invariant lines in the A- and C-phases, 
although the glide and rotation symmetries are spontaneously broken in the B-phase. 
Thus, we obtain some constraints on the winding numbers at ${\bm k}_{sf} =(0, 0)$ and $(0, \pi)$ in the A- and C-phases.

\begin{table}[htbp]
  \begin{tabular}{c|c|c|c|c|c}
% & $\eta$ 
%\\ 
& $\Gamma_z$ & $c(M^{xy},\Gamma_{\rm M})$ & $c(G^{xz},\Gamma_{\rm M})$ & $c(R^{x},\Gamma_{\rm M})$ & $c(T,\Gamma_{\rm M})$ \\
\hline
A-phase & $0$   & -1 & -1 & +1 & -1 \\ \cline{2-2}
        & $\pi$ & -1 & +1 & -1 & -1 \\ \cline{1-2}
C-phase &$0$    & -1 & +1 & -1 & -1 \\ \cline{2-2}
        &$\pi$  & -1 & -1 & +1 & -1 \\ 
 \hline
  \end{tabular}
  \caption{Commutation (anti-commutation) relations of the mirror-chiral operator $\Gamma_{\rm M}$ with 
the crystal symmetry and time-reversal operators are represented by $+1$ ($-1$).  
}
  \label{tab3}
\end{table}

\begin{table}[htbp]
  \begin{tabular}{c|c|c|c|c|c}
% & $\eta$ 
%\\ 
& $\Gamma_z$ & $c(M^{xy},\Gamma_{\rm G})$ & $c(G^{xz},\Gamma_{\rm G})$ & $c(R^{x},\Gamma_{\rm G})$ & $c(T,\Gamma_{\rm G})$ \\
\hline
A-phase & $0$   & +1 & +1 & +1 & +1 \\ \cline{2-2}
        & $\pi$ & -1 & +1 & -1 & -1 \\ \cline{1-2}
C-phase &$0$    & +1 & -1 & -1 & -1 \\ \cline{2-2}
        &$\pi$  & -1 & -1 & +1 & +1 \\
 \hline
  \end{tabular}
  \caption{Commutation (anti-commutation) relations of the glide-chiral operator $\Gamma_{\rm G}$ with 
the crystal symmetry and time-reversal operators.  
}
  \label{tab4}
\end{table}

\begin{table}[htbp]
  \begin{tabular}{c|c|c|c|c|c}
% & $\eta$ 
%\\ 
& $\Gamma_z$ & $c(M^{xy},\Gamma_{\rm R})$ & $c(G^{xz},\Gamma_{\rm R})$ & $c(R^{x},\Gamma_{\rm R})$ & $c(T,\Gamma_{\rm R})$ \\
\hline
A-phase & $0$   & +1 & -1 & -1 & -1 \\ \cline{2-2}
        & $\pi$ & -1 & +1 & -1 & -1 \\ \cline{1-2}
C-phase &$0$    & +1 & +1 & +1 & +1 \\ \cline{2-2}
        &$\pi$  & -1 & -1 & +1 & +1 \\
 \hline
  \end{tabular}
  \caption{Commutation (anti-commutation) relations of the rotation-chiral operator $\Gamma_{\rm R}$ with 
the crystal symmetry and time-reversal operators.  
}
  \label{tab5}
\end{table}

The commutation (anti-commutation) relations between crystal symmetry operators 
$M^{xy}$, $G^{xz}$, $R^x$ and chiral operators $\Gamma_{\rm M}$, $\Gamma_{\rm G}$, and $\Gamma_{\rm R}$ 
are summarized in Tables~\ref{tab3}, \ref{tab4}, and \ref{tab5}. 
From these algebra, we find that only ${\cal \omega}_{\rm G}(0)$ and ${\cal \omega}_{\rm R}^{0}$ may be nontrivial. 
Interestingly, all the winding numbers at $k_z=\pi$ vanish as a consequence of the nonsymmorphic glide symmetry. 
The mirror-winding number at $k_z=0$ also vanishes in both A- and C-phases. 
Furthermore, we see that the rotation-winding number ${\cal \omega}_{\rm R}^{0}$ disappears in the A-phase, 
while the glide-winding number ${\cal \omega}_{\rm G}(0)$ disappears in the C-phase. 
These symmetry constraints are consistent with our numerical calculations summarized in Table.~\ref{tab6}, 
and also consistent with recently obtained general rules for winding numbers~\cite{Xiong}.

\begin{table}[htbp]
{\renewcommand\arraystretch{1.2}
  \begin{tabular}{c|c|c}
% & $\eta$ 
%\\ 
& $|\eta|>1$ & $|\eta| < 1$ \\
\hline
${\cal \omega}_{\rm G}(0)$ & -2 & \textcolor{blue}{0} \\
\hline
${\cal \omega}_{\rm R}^{0}$ & \textcolor{blue}{0} & -2 \\
\hline
  \end{tabular}
}
  \caption{Nontrivial winding numbers of the $\Gamma$-FS.  
The other winding numbers are proved to be zero owing to the symmetry constraints. 
The zeros in the table are also ensured by the adiabatic connection from 
the TRS invariant A- or C-phases. 
}
  \label{tab6}
\end{table}

In addition to the glide-winding number ${\cal \omega}_{\rm G}(0)$ discussed in Sec.~\ref{sec:glide-AIII}, 
we may have a nontrivial rotation-winding number, which is introduced below for completeness.  
Combining the $\pi$-rotation symmetry 
with TRS, we define the magnetic $\pi$-rotation symmetry by $T' = R_\pi^x T = -i s_z \sigma_x K$. 
The BdG Hamiltonian is invariant
\begin{align}
& %\hspace{-7mm}
T'_{\rm BdG} \tilde{H}_{\rm BdG}(\k) T_{\rm BdG}'^{\,\,\,\,\,\,\,\,-1} = \tilde{H}_{\rm BdG}(-k_x,k_y,k_z), 
\end{align}
under the magnetic $\pi$-rotation in the Nambu space, 
\begin{align}
T'_{\rm BdG} &=
\left(
\begin{array}{cc}
T'  & 0 \\
0 & T'^* \\
\end{array}
\right)_{\tau}
= T' \otimes \tau_z, 
\end{align}
not only in the rotation invariant A- and C-phases but also in the B-phase. 
According to the classification by $K$-theory~\cite{Shiozaki2014}, the 2D Hamiltonian of D class on the $k_z = 0$ or $\pi$ plane 
is specified by a $\mathbb{Z} \oplus \mathbb{Z}$ topological invariant by implementing the magnetic $\pi$-rotation symmetry. 
The relations $\left(T'_{\rm BdG}\right)^2 = 1$ and $\left[\,T'_{\rm BdG}, C \,\right] = 0$ are used there. 
One of the integer topological numbers is the rotation-winding number given by Eq.~(\ref{rotation-winding}), 
where the rotation-chiral operator is $\Gamma_{\rm R} = T'_{\rm BdG} C = s_z \sigma_x \tau_y$.

\subsection{Topological transition in B-phase}\label{sec:magnetic-rotation}

In Sec.~\ref{sec:symmetry}, symmetry constraints on the winding numbers have been proved 
in the A- and C-phases. In this subsection the B-phase is discussed. 
We again see that ${\cal \omega}_{\rm G}(\pi) = {\cal \omega}_{\rm R}^{\pi}=0$ owing to the mirror symmetry. 
On the other hand, we obtain ${\cal \omega}_{\rm G}(0) =-2$ when $|\eta|>1$, 
while ${\cal \omega}_{\rm R}^{0}=-2$  when $|\eta|<1$ (see Table~\ref{tab6}). 
The Majorana cone discussed in Sec.~\ref{sec:mirror} is protected by these winding numbers as well.

At $|\eta|=1$ the jump of the winding numbers ${\cal \omega}_{\rm G}(0)$ and ${\cal \omega}_{\rm R}^{0}$ indicates 
the gap closing. Equation~(\ref{subsector-OP_0}) shows that the superconducting gap on the $k_z=0$
plane actually disappears at $|\eta|=1$. 
This gap node has been reported as unusual ``quadratic line node''~\cite{Yanase_UPt3_Weyl}. 
In contrast to the usual linear line node with $\Delta({\bm k}) \propto |k_z|$, which appears in the purely 
$f$-wave $E_{\rm 2u}$-state~\cite{Sauls,Joynt}, the line node of the generic $E_{\rm 2u}$-state is accompanied by 
the quadratic behavior, $\Delta({\bm k}) \propto |k_z|^2$. 
Such an unusual nodal structure at $|\eta|=1$ has been attributed to the pair annihilation of 
Weyl nodes~\cite{Yanase_UPt3_Weyl}. It can also be viewed as a criticality of topological phase transition 
specified by  ${\cal \omega}_{\rm G}(0)$ and ${\cal \omega}_{\rm R}^{0}$.

In contrast to the $k_z=0$ plane, all of the winding numbers on the $k_z=\pi$ plane are zero irrespective of $\eta$. 
Thus, the gap closing enforced by the change of winding numbers does not occur at $k_z=\pi$. 
This is consistent with the numerical result showing the finite superconducting gap on the $k_z=\pi$ plane.

\section{Summary and discussions}

We investigated topologically nontrivial superconducting phases in UPt$_3$. 
Taking into account the FSs reported by first principles band structure calculation 
and quantum oscillation experiments, we have calculated the topological invariants specifying the 
superconducting states and demonstrated topological surface states.

Among a variety of topological properties in UPt$_3$, the most intriguing result is the nontrivial 
glide-$\Z_2$ invariant in the TRS invariant A-phase. By using the $K$-theory for topological nonsymmorphic 
insulators/superconductors, we showed that the glide-$\Z_2$ invariant is the strong topological index 
specifying the 3D glide-even superconductivity of class DIII. 
Although UPt$_3$ is a gapless SC in the bulk, the glide-$\Z_2$ invariant is well-defined and nontrivial. 
Thus, the UPt$_3$ A-phase can be reduced to a 3D gapped TNSC with keeping 
double Majorana cone surface states, when the point nodes are removed by some perturbations. 
By these findings, UPt$_3$ is identified as a 3D gapless TNSC. 
At our best knowledge, this is the first proposal for the material realization of emergent topological 
superconductivity enriched by nonsymmorphic space group symmetry.

Not only the A-phase but also the B- and C-phases have been identified as symmetry-enriched topological superconducting states. 
Combining the crystal symmetries of UPt$_3$ with the TRS and PHS, we find 
topological invariants and surface states as follows. 
\begin{itemize}
\item
Double Majorana cone protected by the glide-$\Z_2$ invariant in the A-phase
\item
Chiral Majorana arcs in the Weyl B-phase
\item
Majorana cone protected by the mirror Chern number in the A-, B-, and C-phases
\item
Majorana flat band protected by the glide-winding number in the A-phase and ``half'' of the B-phase
\end{itemize}
It has been proved that the other mirror Chern number and winding numbers must be trivial because of 
the constraints by symmetry.

From the results obtained in this paper, we notice rich topological properties of superconducting UPt$_3$. 
Underlying origins of such topological superconducting phases are as follows. 
(1) Spin-triplet odd-parity superconductivity, which is often a platform of topological SC. 
(2) 2D $E_{\rm 2u}$ representation, which allows multiple superconducting phases distinguished by symmetry. 
(3) Nonsymmorphic space group symmetry $P6_{3}/mmc$, which gives rise to following features distinct from symmorphic systems, 
\begin{enumerate}
\item 
Classification of topological insulators and SCs changes, and allows emergent topological phases.
\item 
Dirac nodal lines yield the paired FSs which correspond to the pseudospin degree of freedom 
in glide-subsectors. 
\item
The sublattice-singlet $d$-wave pairing naturally admixes with the $f$-wave pairing, and 
leads to the nontrivial glide-$\Z_2$ invariant.
\item
Most mirror Chern numbers and winding numbers are forced to be zero, 
and do not support topological surface states.
\end{enumerate}
Thus, an old heavy fermion superconductor UPt$_3$ is a precious platform of 
topological superconductivity enriched by nonsymmorphic space group symmetry.

\begin{acknowledgments}
The authors are grateful to A. Daido, S. Kobayashi, M. Sato, and S. Sumita for fruitful discussions. 
This work was supported by Grant-in Aid for Scientific Research on Innovative Areas ``J-Physics'' (JP15H05884) 
and ``Topological Materials Science'' (JP16H00991) from JSPS of Japan, and by JSPS KAKENHI Grant Numbers 
JP15K05164 and JP15H05745. 
K.S.\ is supported by JSPS Postdoctoral Fellowship for Research Abroad. 

\end{acknowledgments}

\bibliography{reference}

\begin{thebibliography}{33}%
\makeatletter
\providecommand \@ifxundefined [1]{%
 \@ifx{#1\undefined}
}%
\providecommand \@ifnum [1]{%
 \ifnum #1\expandafter \@firstoftwo
 \else \expandafter \@secondoftwo
 \fi
}%
\providecommand \@ifx [1]{%
 \ifx #1\expandafter \@firstoftwo
 \else \expandafter \@secondoftwo
 \fi
}%
\providecommand \natexlab [1]{#1}%
\providecommand \enquote  [1]{``#1''}%
\providecommand \bibnamefont  [1]{#1}%
\providecommand \bibfnamefont [1]{#1}%
\providecommand \citenamefont [1]{#1}%
\providecommand \href@noop [0]{\@secondoftwo}%
\providecommand \href [0]{\begingroup \@sanitize@url \@href}%
\providecommand \@href[1]{\@@startlink{#1}\@@href}%
\providecommand \@@href[1]{\endgroup#1\@@endlink}%
\providecommand \@sanitize@url [0]{\catcode `\\12\catcode `\$12\catcode
  `\&12\catcode `\#12\catcode `\^12\catcode `\_12\catcode `\%12\relax}%
\providecommand \@@startlink[1]{}%
\providecommand \@@endlink[0]{}%
\providecommand \url  [0]{\begingroup\@sanitize@url \@url }%
\providecommand \@url [1]{\endgroup\@href {#1}{\urlprefix }}%
\providecommand \urlprefix  [0]{URL }%
\providecommand \Eprint [0]{\href }%
\providecommand \doibase [0]{http://dx.doi.org/}%
\providecommand \selectlanguage [0]{\@gobble}%
\providecommand \bibinfo  [0]{\@secondoftwo}%
\providecommand \bibfield  [0]{\@secondoftwo}%
\providecommand \translation [1]{[#1]}%
\providecommand \BibitemOpen [0]{}%
\providecommand \bibitemStop [0]{}%
\providecommand \bibitemNoStop [0]{.\EOS\space}%
\providecommand \EOS [0]{\spacefactor3000\relax}%
\providecommand \BibitemShut  [1]{\csname bibitem#1\endcsname}%
\let\auto@bib@innerbib\@empty
%</preamble>



\bibitem{Qi-Zhang}
X.-L. Qi and S.-C. Zhang, 
Topological insulators and superconductors, 
Rev. Mod. Phys. {\bf 83}, 1057 (2011).

\bibitem{Tanaka_review}
Y. Tanaka, M. Sato, and N. Nagaosa, 
Symmetry and topology in superconductors -odd-frequency pairing and edge states-, 
J. Phys. Soc. Jpn. {\bf 81}, 011013 (2012).

\bibitem{Sato-Fujimoto_review}
M. Sato and S. Fujimoto, 
Majorana Fermions and Topology in Superconductors, 
J. Phys. Soc. Jpn. {\bf 85}, 072001 (2016). 

\bibitem{Lutchyn2010}
R. M. Lutchyn, J. D. Sau, and S. Das Sarma, 
Majorana Fermions and a Topological Phase Transition in Semiconductor-Superconductor Heterostructures, 
Phys. Rev. Lett. {\bf 105}, 077001 (2010).

\bibitem{Mourik}
V. Mourik, K. Zuo, S. M. Frolov, S. R. Plissard, E. P. A. M. Bakkers, and L. P. Kouwenhoven, 
Signatures of Majorana Fermions in Hybrid Superconductor-Semiconductor Nanowire Devices, 
Science {\bf 336}, 1003 (2012).

\bibitem{Nadj-Perge}
S. Nadj-Perge, I. K. Drozdov, J. Li, H. Chen, S. Jeon, J. Seo, A. H. MacDonald, B. A. Bernevig, 
and A. Yazdani, 
Observation of Majorana fermions in ferromagnetic atomic chains on a superconductor, 
Science {\bf 346}, 602 (2014). 

\bibitem{Fu-Kane2008}
L. Fu and C. L. Kane, 
Superconducting Proximity Effect and Majorana Fermions at the Surface of a Topological Insulator, 
Phys. Rev. Lett. {\bf 100}, 096407 (2008).

\bibitem{Sun2016}
H.-H. Sun, K.-W. Zhang, L.-H. Hu, C. Li, G.-Y. Wang, H.-Y. Ma, Z.-A. Xu, C.-L. Gao, D.-D. Guan, 
Y.-Y. Li, C. Liu, D. Qian, Y. Zhou, L. Fu, S.-C. Li, F.-C. Zhang, and J.-F. Jia, 
Majorana Zero Mode Detected with Spin Selective Andreev Reflection in the Vortex of 
a Topological Superconductor, 
Phys. Rev. Lett. {\bf 116}, 257003 (2016).


\bibitem{Read-Green}
N. Read and D. Green, 
Paired states of fermions in two dimensions with breaking of parity and time-reversal symmetries 
and the fractional quantum Hall effect, 
Phys. Rev. B {\bf 61}, 10267 (2000).


\bibitem{Kitaev2001}
A. Y. Kitaev, 
Unpaired Majorana fermions in quantum wires, 
Phys. Usp. {\bf 44}, 131 (2001).

\bibitem{Schnyder}
A. P. Schnyder, S. Ryu, A. Furusaki, and A. W. W. Ludwig, 
Classification of topological insulators and superconductors in three spatial dimensions, 
Phys. Rev. B {\bf 78}, 195125 (2008).

\bibitem{Sato2010}
M. Sato, 
Topological odd-parity superconductors, 
Phys. Rev. B {\bf 81}, 220504 (2010).


\bibitem{Kasahara2007}
Y. Kasahara, T. Iwasawa, H. Shishido, T. Shibauchi, K. Behnia, Y. Haga, 
T. D. Matsuda, Y. Onuki, M. Sigrist, and Y. Matsuda, 
Exotic superconducting properties in the electron-hole-compensated 
heavy-fermion gsemimetalh URu$_2$Si$_2$, 
Phys. Rev. Lett. {\bf 99}, 116402 (2007).

\bibitem{Yano2008}
K. Yano, T. Sakakibara, T. Tayama, M. Yokoyama, H. Amitsuka, Y. Homma, 
P. Miranovi\'c, M. Ichioka, Y. Tsutsumi, and K. Machida, 
Field-angle-dependent specific heat measurements and gap determination of 
a heavy fermion superconductor URu$_2$Si$_2$, 
Phys. Rev. Lett. {\bf 100}, 017004 (2008).

\bibitem{Kittaka2016}
S. Kittaka, Y. Shimizu, T. Sakakibara, Y. Haga, E. Yamamoto, Y. \=Onuki, 
Y. Tsutsumi, T. Nomoto, H. Ikeda, and K. Machida, 
Evidence for chiral $d$-Wave Superconductivity in URu$_2$Si$_2$ from 
the field-angle variation of its specific heat, 
J. Phys. Soc. Jpn. {\bf 85}, 033704 (2016). 


\bibitem{BiswasSrPtAs}
P. K. Biswas, H. Luetkens, T. Neupert, T. Sturzer, C. Baines, G. Pascua, A. P. Schnyder, 
M. H. Fischer, J. Goryo, M. R. Lees, H. Maeter, F. Bruckner, H.-H. Klauss, M. Nicklas, 
P. J. Baker, A. D. Hillier, M. Sigrist, A. Amato, and D. Johrendt, 
Evidence for superconductivity with broken time-reversal symmetry in locally noncentrosymmetric SrPtAs, 
Phys. Rev. B {\bf 87}, 180503(R) (2013).

\bibitem{Sr2RuO4_review2012}
Y. Maeno, S. Kittaka, T. Nomura, S. Yonezawa, and K. Ishida, 
Evaluation of Spin-Triplet Superconductivity in Sr$_2$RuO$_4$, 
J. Phys. Soc. Jpn. {\bf 81}, 011009 (2012).

\bibitem{Sasaki2011}
S. Sasaki, M. Kriener, K. Segawa, K. Yada, Y. Tanaka, M. Sato, and Y. Ando, 
Topological Superconductivity in Cu$_x$Bi$_2$Se$_3$, 
Phys. Rev. Lett. {\bf 107}, 217001 (2011).

\bibitem{Aoki_review}
D. Aoki and J. Flouquet, 
Ferromagnetism and Superconductivity in Uranium Compounds, 
J. Phys. Soc. Jpn. {\bf 81}, 011003 (2012).


\bibitem{Stewart}
G. R. Stewart, Z. Fisk, J. O. Willis, and J. L. Smith, 
Possibility of coexistence of bulk superconductivity and spin fluctuations in UPt$_3$, 
Phys. Rev. Lett. {\bf 52}, 679 (1984). 


\bibitem{Fisher}
R. A. Fisher, S. Kim, B. F. Woodfield, N. E. Phillips, L. Taillefer, K. Hasselbach, 
J. Flouquet, A. L. Giorgi, and J. L. Smith, 
Specific heat of UPt$_3$: Evidence for unconventional superconductivity, 
Phys. Rev. Lett. {\bf 62}, 1411 (1989). 

\bibitem{Bruls}
G. Bruls, D. Weber, B. Wolf, P. Thalmeier, B. Luthi, A. de Visser, and A. Menovsky, 
Strain-order-parameter coupling and phase diagrams in superconducting UPt$_3$, 
Phys. Rev. Lett. {\bf 65}, 2294 (1990).

\bibitem{Adenwalla}
S. Adenwalla, S. W. Lin, Q. Z. Ran, Z. Zhao, J. B. Ketterson, J. A. Sauls, L. Taillefer, 
D. G. Hinks, M. Levy, and Bimal K. Sarma, 
Phase diagram of UPt$_3$ from ultrasonic velocity measurements,
Phys. Rev. Lett. \textbf{65}, 2298 (1990). 
%

\bibitem{Tou_UPt3}
H. Tou, Y. Kitaoka, K. Ishida, K. Asayama, N. Kimura, Y. Onuki, 
E. Yamamoto, Y. Haga, and K. Maezawa, 
Nonunitary spin-triplet superconductivity in UPt$_3$: 
evidence from $^{195}$Pt Knight shift study, 
Phys. Rev. Lett. \textbf{80}, 3129 (1998). 




\bibitem{Sigrist-Ueda}
M.~Sigrist and K.~Ueda, 
Phenomenological theory of unconventional superconductivity, 
Rev. Mod. Phys. {\bfseries 63}, 239 (1991).

\bibitem{Sauls}
J. A. Sauls, 
The order parameter for the superconducting phases of UPt$_3$, 
Adv. Phys. {\bf 43}, 113 (1994).

\bibitem{Joynt}
R. Joynt and L. Taillefer, 
The superconducting phases of UPt$_3$, 
Rev. Mod. Phys. {\bf 74}, 235 (2002).

\bibitem{Strand}
J. D. Strand, D. J. Bahr, D. J. Van Harlingen, J. P. Davis, W. J. Gannon, 
and W. P. Halperin, 
The transition between real and complex superconducting order parameter phases in UPt$_3$,
Science {\bf 328}, 1368 (2010).

\bibitem{Schemm}
E. R. Schemm, W. J. Gannon, C. M. Wishne, W. P. Halperin, and A. Kapitulnik, 
Observation of broken time-reversal symmetry in the heavy-fermion superconductor UPt$_3$, 
Science \textbf{345}, 190 (2014).  


\bibitem{Comment2}
Symmetry breaking by a weak crystal distortion has been reported~\cite{Walko},
although its reliability is under debate. We here assume high symmetry 
space group $P6_{3}/mmc$ and examine the effect of broken glide symmetry 
in Sec.~\ref{sec:broken_glide}. 

\bibitem{Walko}
D. A. Walko, J.-I. Hong, T. V. Chandrasekhar Rao, Z. Wawrzak, D. N. Seidman, 
W. P. Halperin, and M. J. Bedzyk, 
Crystal structure assignment for the heavy-fermion superconductor UPt$_3$, 
Phys. Rev. B {\bf 63}, 054522 (2001). 


\bibitem{Norman1995}
M. R. Norman, 
Odd parity and line nodes in heavy-fermion superconductors, 
Phys. Rev. B {\bf 52}, 15093 (1995).  


\bibitem{Blount}
E. I. Blount, 
Symmetry properties of triplet superconductors, 
Phys. Rev. B {\bf 32}, 2935 (1985). 

\bibitem{Kobayashi-Sato}
S. Kobayashi, K. Shiozaki, Y. Tanaka, and M. Sato,
Topological Blount's theorem of odd-parity superconductors, 
Phys. Rev. B {\bf 90}, 024516 (2014). 

\bibitem{Micklitz2009}
T. Micklitz and M. R. Norman, 
Odd parity and line nodes in nonsymmorphic superconductors, 
Phys. Rev. B {\bf 80}, 100506(R) (2009). 

\bibitem{Micklitz2017-1}
T. Micklitz and M. R. Norman, 
Nodal lines and nodal loops in nonsymmorphic odd-parity superconductors, 
Phys. Rev. B {\bf 95}, 024508 (2017).

\bibitem{Yanase_UPt3_Weyl}
Y. Yanase, 
Nonsymmorphic Weyl superconductivity in UPt$_3$ based on E$_{2u}$ representation, 
Phys. Rev. B {\bf 94}, 174502 (2016). 

\bibitem{Kobayashi-Yanase-Sato}
S. Kobayashi, Y. Yanase, and M. Sato, 
Topologically stable gapless phases in nonsymmorphic superconductors, 
Phys. Rev. B {\bf 94}, 134512 (2016). 

\bibitem{Nomoto}
T. Nomoto and H. Ikeda, 
Exotic Multigap Structure in UPt$_3$ Unveiled by a First-Principles Analysis, 
Phys. Rev. Lett. {\bf 117}, 217002 (2016). 



\bibitem{Fang-Fu2015}
C. Fang and L. Fu, 
New classes of three-dimensional topological crystalline insulators: Nonsymmorphic and magnetic, 
Phys. Rev. B {\bf 91}, 161105 (2015).

\bibitem{Shiozaki2015}
K. Shiozaki, M. Sato, and K. Gomi, 
$Z_2$ topology in nonsymmorphic crystalline insulators: 
M\"obius twist in surface states
Phys. Rev. B {\bf 91}, 155120 (2015).

\bibitem{Shiozaki2016}
K. Shiozaki, M. Sato, and K. Gomi, 
Topology of nonsymmorphic crystalline insulators and superconductors, 
Phys. Rev. B {\bf 93}, 195413 (2016). 

\bibitem{Varjas2015}
D. Varjas, F. de Juan, and Y.-M. Lu, 
Bulk invariants and topological response in insulators and superconductors with nonsymmorphic symmetries, 
Phys. Rev. B {\bf 92}, 195116 (2015). 

\bibitem{Liu2016}
Q.-Z. Wang and C.-X. Liu, 
Topological nonsymmorphic crystalline superconductors, 
Phys. Rev. B {\bf 93}, 020505(R) (2016). 

\bibitem{Kitaev2009}
A. Kitaev, 
Periodic table for topological insulators and superconductors, 
AIP Conf. Proc. No.~1134 (AIP, New York, 2009), p.~22. 


\bibitem{Ryu2010}
S. Ryu, A. P. Schnyder, A. Furusaki, and A. W. W. Ludwig, 
Topological insulators and superconductors: 
ten-fold way and dimensional hierarchy, 
New J. Phys. {\bf 12}, 065010 (2010).


\bibitem{Morimoto2013}
T. Morimoto and A. Furusaki, 
Topological classification with additional symmetries from Clifford algebras, 
Phys. Rev. B {\bf 88}, 125129 (2013). 

\bibitem{Shiozaki2014}
K. Shiozaki and M. Sato, 
Topology of crystalline insulators and superconductors, 
Phys. Rev. B {\bf 90}, 165114 (2014). 

\bibitem{KHgX}
Z. Wang, A. Alexandradinata, R. J. Cava and B. A. Bernevig, 
Hourglass fermions, 
Nature {\bf 532}, 189 (2016). 

\bibitem{KHgX_ARPES}
J. Ma, C. Yi, B. Lv, Z. Wang, S. Nie, L. Wang, L. Kong, Y. Huang, P. Richard, P. Zhang, 
K. Yaji, K. Kuroda, S. Shin, H. Weng, B. A. Bernevig, Y. Shi, T. Qian, H. Ding, 
Experimental evidence of hourglass fermion in the candidate nonsymmorphic topological insulator KHgSb, 
Sci. Adv. {\bf 3}, e1602415 (2017). 

\bibitem{CeNiSn}
P. Y. Chang, O. Erten, and P. Coleman, 
M\" obius Kondo Insulators, 
Nat. Phys. online (2017), doi:10.1038/nphys4092. 
%arXiv:1603.03435. 


\bibitem{Taillefer1988}
L. Taillefer and G. G. Lonzarich, 
Heavy-fermion quasiparticles in UPt$_3$, 
Phys. Rev. Lett. {\bf 60}, 1570 (1988); 
M. R. Norman, R. C. Albers, A. M. Boring, and N. E. Christensen, 
Fermi surface and effective masses for the heavy-electron superconductors UPt$_3$, 
Solid State Commun. {\bf 68}, 245 (1988). 


\bibitem{Kimura_UPt3}
N. Kimura, R. Settai, Y. Onuki, H. Toshima, E. Yamamoto, K. Maezawa, H. Aoki, and H. Harima, 
Magnetoresistance and de Haas-van Alphen effect in UPt$_3$, 
J. Phys. Soc. Jpn. {\bf 64}, 3881 (1995). 


\bibitem{McMullan}
G. J. McMullan, P. M. C. Rourke, M. R. Norman, A. D. Huxley, N. Doiron-Leyraud, 
J. Flouquet, G. G. Lonzarich, A. McCollam, and S. R. Julian, 
The Fermi surface and f-valence electron count of UPt$_3$, 
New J. Phys, {\bf 10}, 053029 (2008).


\bibitem{Kane-Mele}
C. L. Kane and E. J. Mele, 
Quantum Spin Hall Effect in Graphene, 
Phys. Rev. Lett. {\bf 95}, 226801 (2005). 

\bibitem{Saito_MoS2}
Y. Saito, Y. Nakamura, M. S. Bahramy, Y. Kohama, J. Ye, Y. Kasahara, 
Y. Nakagawa, M. Onga, M. Tokunaga, T. Nojima, Y. Yanase, Y. Iwasa,
Superconductivity protected by spin-valley locking in gate-tuned MoS$_2$, 
Nat. Phys. {\bf 12}, 144 (2016). 
%doi:10.1038/nphys3580. 

\bibitem{Fischer}
M.~H. Fischer, F.~Loder, and M.~Sigrist, 
Superconductivity and local noncentrosymmetricity in crystal lattices, 
Phys. Rev. B {\bfseries 84}, 184533 (2011).
%
\bibitem{JPSJ.81.034702}
D.~Maruyama, M.~Sigrist, and Y.~Yanase,
Locally non-centrosymmetric superconductivity in multilayer systems, 
J. Phys. Soc. Jpn. {\bfseries 81}, 034702 (2012).
%

\bibitem{Burkov-Hook-Balents2011}
A. A. Burkov, M. D. Hook, and Leon Balents, 
Topological nodal semimetals, 
Phys. Rev. B {\bf 84}, 235126 (2011). 

\bibitem{Bradley}
C. J. Bradley and A. P. Cracknell, 
{\it The Mathematical Theory of Symmetry in Solids} 
(Oxford University Press, Oxford, 1972).

\bibitem{Young2012}
S. M. Young, S. Zaheer, J. C. Y. Teo, C. L. Kane, E. J. Mele, and A. M. Rappe, 
Dirac Semimetal in Three Dimensions, 
Phys. Rev. Lett. {\bf 108}, 140405 (2012). 

\bibitem{Watanabe-Po-Vishwanath}
%H. Watanabe, H. C. Po, A. Vishwanath, and M. Zaletel,
%Proc. Natl. Acad. Sci. {\bf 112}, 14551 (2015).
H. Watanabe, H. C. Po, M. P. Zaletel, and A. Vishwanath, 
Filling-Enforced Gaplessness in Band Structures of the 230 Space Groups, 
Phys. Rev. Lett. {\bf 117}, 096404 (2016). 

\bibitem{Niu2016}
Q. Niu, W. C. Yu, K. Y. Yip, Z. L. Lim, H. Kotegawa, E. Matsuoka, H. Sugawara, 
H. Tou, Y. Yanase, and S. K. Goh, 
Quasilinear quantum magnetoresistance in pressure-induced nonsymmorphic 
superconductor CrAs, 
arXiv:1612.07480. 

\bibitem{Yang2017}
B.-J. Yang, T. A. Bojesen, T. Morimoto, and A. Furusaki, 
Topological semimetals protected by off-centered symmetries in nonsymmorphic crystals, 
Phys. Rev. B {\bf 95}, 075135 (2017). 

\bibitem{Aeppli}
G. Aeppli, E. Bucher, C. Broholm, J. K. Kjems, J. Baumann and J. Hufnagl, 
Magnetic order and fluctuations in superconducting UPt$_3$, 
Phys. Rev. Lett. \textbf{60}, 615 (1988). 

\bibitem{Hayden}
S. M. Hayden, L. Taillefer, C. Vettier, and J. Flouquet, 
Antiferromagnetic order in UPt$_3$ under pressure: evidence for a direct coupling 
to superconductivity, 
Phys. Rev. B \textbf{46}, 8675(R) (1992). 


\bibitem{Machida-Izawa}
Y. Machida, A. Itoh, Y. So, K. Izawa, Y. Haga, E. Yamamoto, N. Kimura,
Y. Onuki, Y. Tsutsumi, and K. Machida, 
Twofold spontaneous symmetry breaking in the heavy-fermion superconductor UPt$_3$, 
Phys. Rev. Lett. \textbf{108}, 157002 (2012).

\bibitem{Tsutsumi2012}
Y. Tsutsumi, K. Machida, Tetsuo Ohmi, and Masa-aki Ozaki,
A Spin Triplet Superconductor UPt$_3$,
J. Phys. Soc. Jpn. {\bf 81}, 074717 (2012). 


\bibitem{Maruyama-Yanase2015}
D. Maruyama and Y. Yanase, 
Electron Correlation Effects in Non-centrosymmetric Metals 
in the Weak Coupling Regime, 
J. Phys. Soc. Jpn. {\bf 84}, 074702 (2015). 


\bibitem{Shiozaki-Sato-Gomi2017}
K. Shiozaki, M. Sato, and K. Gomi, 
Topological Crystalline Materials 
-- General Formulation, Module Structure, and Wallpaper Groups, 
arXiv:1701.08725. 

\bibitem{Shiozaki-Yanase2017}
K. Shiozaki and Y. Yanase, in preparation. 


\bibitem{Goswami}
P. Goswami and A. H. Nevidomskyy, 
Topological Weyl superconductor to diffusive thermal Hall metal crossover 
in the B phase of UPt$_3$, 
Phys. Rev. B {\bf 92}, 214504 (2015). 



\bibitem{Murakami}
S. Murakami, 
Phase transition between the quantum spin Hall and insulator phases in 3D: 
emergence of a topological gapless phase, 
New J. Phys. {\bf 9}, 356 (2007). 
\bibitem{Wan-Vishwanath}
X. Wan, A. M. Turner, A. Vishwanath, and S. Y. Savrasov, 
Topological semimetal and Fermi-arc surface states 
in the electronic structure of pyrochlore Iridates, 
Phys. Rev. B {\bf 83}, 205101 (2011).
\bibitem{Burkov-Balents}
A. A. Burkov and L. Balents, 
Weyl semimetal in a topological insulator multilayer, 
Phys. Rev. Lett. {\bf 107}, 127205 (2011). 
\bibitem{Xu}
S.-Y. Xu, I. Belopolski, N. Alidoust, M. Neupane, C. Zhang, R. Sankar, 
S.-M. Huang, C.-C. Lee, G. Chang, B. Wang, G. Bian, H. Zheng, 
D. S. Sanchez, F. Chou, H. Lin, S. Jia, and M. Z. Hasan, 
Discovery of a Weyl fermion semimetal and topological Fermi arcs, 
Science {\bf 349}, 613 (2015). 
\bibitem{Lv}
B. Q. Lv, N. Xu, H. M. Weng, J. Z. Ma, P. Richard, X. C. Huang, 
L. X. Zhao, G. F. Chen, C. E. Matt, F. Bisti, V. N. Strocov, J. Mesot, 
Z. Fang, X. Dai, T. Qian, M. Shi, and H. Ding, 
Observation of Weyl nodes in TaAs, 
Nat. Phys. {\bf 11}, 724 (2015). 
%\bibitem{Lv2}
%B. Q. Lv, H. M. Weng, B. B. Fu, X. P. Wang, H. Miao, J. Ma, P. Richard, 
%X. C. Huang, L. X. Zhao, G. F. Chen, Z. Fang, X. Dai, T. Qian, and H. Ding,
%Phys. Rev. X {\bf 5}, 031013 (2015). 
\bibitem{Yang}
L. X. Yang, Z. K. Liu, Y. Sun, H. Peng, H. F. Yang, T. Zhang, B. Zhou, 
Y. Zhang, Y. F. Guo, M. Rahn, D. Prabhakaran, Z. Hussain, S.-K. Mo, 
C. Felser, B. Yan, and Y. L. Chen, 
Weyl semimetal phase in the non-centrosymmetric compound TaAs, 
Nat. Phys. {\bf 11}, 728 (2015). 
\bibitem{Huang}
X. Huang, L. Zhao, Y. Long, P. Wang, D. Chen, Z. Yang, H. Liang, 
M. Xue, H. Weng, Z. Fang, X. Dai, and G. Chen, 
Observation of the chiral-anomaly-induced negative magnetoresistance 
in 3D Weyl semimetal TaAs, 
Phys. Rev. X {\bf 5}, 031023 (2015). 


\bibitem{Yamashita2015}
T. Yamashita, Y. Shimoyama, Y. Haga, T. D. Matsuda, E. Yamamoto, Y. Onuki, H. Sumiyoshi, 
S. Fujimoto, A. Levchenko, T. Shibauchi, and Y. Matsuda, 
Colossal thermomagnetic response in the exotic superconductor URu$_2$Si$_2$, 
Nat. Phys. {\bf 11}, 17 (2015); 
H. Sumiyoshi and S. Fujimoto, 
Giant Nernst and Hall effects due to chiral superconducting fluctuations,
Phys. Rev. B {\bf 90}, 184518 (2014).


\bibitem{Thouless}
D. J. Thouless, M. Kohmoto, M. P. Nightingale, and M. den Nijs, 
Quantized Hall conductance in a two-dimensional periodic potential, 
Phys. Rev. Lett. {\bf 49}, 405 (1982).

\bibitem{Kohmoto}
M. Kohmoto, 
Topological invariant and the quantization of the Hall conductance, 
Ann. Phys. {\bf 160}, 343 (1985).

\bibitem{Fukui}
T. Fukui, Y. Hatsugai, and H. Suzuki, 
Chern numbers in discretized Brillouin zone: efficient method of computing 
(spin) Hall conductances, 
J. Phys. Soc. Jpn. {\bf 74}, 1674 (2005).


\bibitem{Ueno-Sato}
Y. Ueno, A. Yamakage, Y. Tanaka, and M. Sato, 
Symmetry-protected Majorana fermions in topological crystalline superconductors: 
theory and application to Sr$_2$RuO$_4$, 
Phys. Rev. Lett. {\bf 111}, 087002 (2013). 

\bibitem{Yoshida2015}
T. Yoshida, M. Sigrist, and Y. Yanase, 
Topological crystalline superconductivity in locally non-centrosymmetric 
multilayer superconductors, 
Phys. Rev. Lett. {\bf 115}, 027001 (2015). 


\bibitem{Comment1}
In the C-phase, the chiral symmetry is not preserved in the glide-subsectors, 
since $\left\{ \Gamma, G^{xz}_{\rm BdG}(k_z)\right\}=0$. 



\bibitem{Tsutsumi_UPt3}
Y. Tsutsumi, M. Ishikawa, T. Kawakami, T. Mizushima, M. Sato, 
M. Ichioka, and K. Machida, 
UPt$_3$ as a topological crystalline superconductor, 
J. Phy. Soc. Jpn. {\bf 82}, 113707 (2013).

\bibitem{Mizushima_3He}
T. Mizushima, M. Sato, and K. Machida, 
Symmetry protected topological order and spin susceptibility 
in superfluid $^3$He-B, 
Phys. Rev. Lett. {\bf 109}, 165301 (2012).

\bibitem{Xiong}
Y. Xiong, A. Yamakage, S. Kobayashi, M. Sato, Y. Tanaka, 
Anisotropic magnetic responses of topological crystalline superconductors, 
Crystals {\bf 7}, 58 (2017).



\end{thebibliography}%

\end{document}